\documentclass{aa} 
\usepackage{graphicx}
\usepackage{natbib}
\bibpunct{(}{)}{;}{a}{}{,} %

\usepackage{amsmath}
\usepackage{amssymb}
\usepackage{wasysym}
\usepackage{appendix}

\usepackage{setspace}
\newcommand{\hs}{\hspace{0.05cm}}
\title{Modeled flux and polarisation signals of horizontally
  inhomogeneous exoplanets, applied to Earth--like planets}

\author{T. Karalidi\inst{1,2} \and D. M. Stam\inst{1}}
        
\institute{ SRON - Netherlands Institute for Space Research,
  Sorbonnelaan 2, 3584 CA, Utrecht, the Netherlands \and Leiden
  Observatory, P.O.\ Box 9513, NL-2300 RA, Leiden, the Netherlands}

\date{Received 28-03-2012 / Accepted 30-08-2012}

\begin{document}

%%%%%%%%%%%%%%%%%%%%%%%%%%%%%%%%%%%%%%%%%%%%%%%%%%%%%%%%%%%%%%%%%%%%%%%%%%
\abstract{}{We present modeled flux and linear polarisation signals of
  starlight that is reflected by spatially unresolved, horizontally
  inhomogeneous planets and discuss the effects of including
  horizontal inhomogeneities on the flux and polarisation signals of
  Earth-like exoplanets.}  {Our code is based on an efficient
  adding--doubling algorithm, which fully includes multiple scattering
  by gases and aerosol/cloud particles. We divide a model planet into
  pixels that are small enough for the local properties of the
  atmosphere and surface (if present) to be horizontally homogeneous.
  Given a planetary phase angle, we sum up the reflected total and
  linearly polarised fluxes across the illuminated and visible part of
  the planetary disk, taking care to properly rotate the polarized
  flux vectors towards the same reference plane.}  {We compared flux
  and polarisation signals of simple horizontally inhomogeneous model
  planets against results of the weighted sum approximation, in which
  signals of horizontally homogeneous planets are combined. Apart from
  cases in which the planet has only a minor inhomogeneity, the
  signals differ significantly. In particular, the shape of the
  polarisation phase function appears to be sensitive to the
  horizontal inhomogeneities.  The same holds true for Earth-like
  model planets with patchy clouds above an ocean and a sandy
  continent. Our simulations clearly show that horizontal
  inhomogeneities leave different traces in flux and polarisation
  signals. Combining flux with polarisation measurements would help
  retrieving the atmospheric and surface patterns on a planet.  }{}

%%%%%%%%%%%%%%%%%%%%%%%%%%%%%%%%%%%%%%%%%%%%%%%%%%%%%%%%%%%%%%%%%%%%%%%%%%

\titlerunning{Horizontally inhomogeneous exoplanets}
\authorrunning{T. Karalidi, and D. M. Stam}

\maketitle

%%%%%%%%%%%%%%%%%%%%%%%%%%%%%%%%%%%%%%%%%%%%%%%%%%%%%%%%%%%%%%%%%%%%%%%%%%
\section{Introduction}

Since \citet{mayorqueloz95} discovered the first planet orbiting
another main sequence star almost two decades ago, the rapid
improvement of detection methods and instruments has yielded hundreds
of exoplanets, including several tens of so--called super--Earths
\citep[see e.g.][]{leger11,charbonneau09,millerricci10,beaulieu06},
and many more will follow in the coming years. The next step in
exoplanet research is the characterisation of the atmospheres and
surfaces (if present) of detected exoplanets: what is their
composition and structure?

Currently, exoplanet atmospheres are being characterised using the
{\em transit method} \citep[see e.g.][]{beaulieu10,millerricci10}.
This method is based on measurements of the wavelength dependendence
of starlight that filters through the upper planetary atmosphere
during the primary transit, or of the planetary flux just before or
after the secondary eclipse. The transit method is mostly applied to
gaseous planets that orbit close to their star. The chances to catch
gaseous planets in wide orbits, such as Jupiter and Saturn in our
Solar System, transiting their star are extremely small, because their
orbital plane should be perfectly aligned with our line of sight and
because their transits are very rare. Earth-sized exoplanets in the
habitable zone of a solar--type star are probably too small and
transit too seldom to reach a sufficient signal-to-noise ratio to do
transit spectroscopy \citep[][]{kaltenegger09}.

A promising method to characterise atmospheres and surfaces (if
present) of exoplanets that are small and/or in wide orbits, is
\textit{direct detection}, in which the starlight that a planet
reflects and/or the thermal radiation that a planet emits is measured
separately from the stellar light (except for some background
starlight). Some instruments that are being designed for such direct
detections are SPHERE (for the VLT) and EPICS (for the European
Extremely Large Telescope, or E-ELT). An example for a space telescope
for direct detection is the New Worlds Observer (NWO)
\citep[][]{cash10}, which is under study by NASA. Through direct
detections, broadband images and/or spectra of various types of
exoplanets will become available in the near future.

Knowing the Solar System planets, it is to be expected that exoplanets
that will be observed are horizontally inhomogeneous, e.g. with clouds
and hazes in patches, such as on Earth and Mars, or in banded
structures, such as, for example, on Jupiter and Saturn. And there
will undoubtedly be solid exoplanets with significant local variations
in surface reflection properties and texture, such as the Earth with
its continents and oceans. Although the lack of spatial resolution in
near future exoplanet observations will merge all spatial variation
into a single image pixel and/or spectrum, accounting for the
existence of horizontal inhomogeneities will be important when trying
to retrieve planet characteristics. For example, efforts to identify
spectral signatures of life on other planets will face various
challenges, such as clouds masking or mimicking the signatures of
vegetation \citep[][and references therein]{tinetti06b}. The cloud
coverage will also influence the retrieval of mixing ratios of
atmospheric gases, such as water vapour and oxygen, from reflected
light spectra. In particular, the larger the fraction of clouds across
an exoplanetery surface, the smaller the depth of gaseous absorption
bands. Absorption band depths are, however, also influenced by the
cloud top altitudes, with higher clouds yielding shallower absorption
bands \citep[for examples, see][]{stam08}. The distribution of cloud
top altitudes across a planet will thus also be a parameter to take
into account.

Horizontal inhomogenities can have large effects on the flux of
starlight that is reflected by a planet. In particular, \citet{ford01}
have shown that an Earth analogue planet without clouds would show
diurnal flux variations of up to $150$\% due to the variation of the
albedo of different regions on the planet. In full agreement,
\citet{oakley09} have calculated that in absence of an atmosphere, the
flux of an Earth-like exoplanet would show a clear diurnal variability
as different continents would rotate in and out of the field of view
of an observer. The presence of clouds in the Earth's atmosphere
significantly complicates the characterisation of various surface
types. \citet{ford01} show, for example, that an Earth-like cloud
pattern would suppress the diurnal flux variations to as little as
20$\%$, while \citet{oakley09} indicates that characterizing the
surface of the planet is possible only for cloud coverages
significantly lower than the average coverage on Earth ($\lesssim$
25\% versus $\sim$ 60\%). To quantitatively estimate the effects of
horizontal inhomogeneities due to clouds or surface features on
observed spectra, and to be able to account for such variations in the
retrieval of planet characteristics from future observations,
numerical codes are essential tools.

In this paper, we present our numerical code to calculate spectra of
starlight that is reflected by spatially unresolved, horizontally
inhomogeneous exoplanets. The main difference with other codes for
horizontally inhomogeneous planets \citep[such as those used by]
[]{ford01,oakley09,tinetti06b} is that it can be used to calculate not
only the flux of reflected starlight but also its state (degree and
direction) of polarisation.  Polarimetry promises to play an important
role in exoplanet research both for exoplanet detection and
characterisation. In particular, because the direct starlight is
unpolarized, while the starlight that is reflected by a planet will
usually be polarized \citep[see e.~g.~][]{zugger10, stam06,
  stamhovenier04, saar03, seager00}, polarimetry can increase the
planet--to--star contrast ratio by 3 to 4 orders of magnitude
\citep{keller10}, thus facilitating the detection of an exoplanet that
might otherwise be lost in the glare of its parent star. Polarimetry
will not only help to detect a planet, it will also confirm the status
of the object, since background objects will usually be unpolarized.

The importance of polarimetry for studying planetary atmospheres and
surfaces has been shown many times using observations of the Earth and
other Solar System planets \citep[see for example][]{hansenhovenier74,
  hansentravis74, mishchenko90, tomasko09}, as well as by modeling of
Solar System planets or giant exoplanets
\citep[e.~g.~][]{madhusudhan12, stam08, stam03, stamhovenier04,
  saar03, seager00}. In particular the sensitivity of polarisation to
the microphysical properties of the scatterers in the planetary
atmosphere, make it a crucial tool for braking degeneracies that flux
only observations can have.

The radiative transfer calculations in our code are based on an
efficient adding-doubling algorithm \citep[][]{dehaan87} which fully
includes multiple scattering by gases and aerosol/cloud particles,
that was used before for flux and polarisation calculations for
gaseous and terrestrial exoplanets by \citet{stam03, stamhovenier04,
  stam08, karalidi11}. These authors, however, assumed each exoplanet
to be horizontally homogeneous, such that it could be treated as a
single starlight scattering 'particle', which allowed for a very fast
integration of the reflected flux and polarisation signals across the
planet's disk for the whole planetary phase angle range
\citep[see][for a description of this disk-integration
  algorithm]{stam06}. With this horizontally homogeneous code, the
signals of horizontally inhomogeneous planets can be simulated using
the so-called weighted sum approximation: signals of homogeneous
planets are multiplied by a weighting factor and summed to yield the
final signal.  We, on the other hand, divide a horizontally
inhomogeneous model planet into pixels that are small enough for the
local properties of the atmosphere and surface (if present) to be
horizontally homogeneous. For each type of pixel, we perform
adding-doubling radiative transfer calculations \citep[][]{dehaan87}
and, given the planetary phase angle, we sum up the reflected total
and polarised fluxes across the illuminated and visible part of the
planetary disk. Our code for horizontally inhomogeneous planets allows
investiging the applicability of the weighted sum approximation, and
the effects of horizontal inhomogeneities on the flux and in
particular the polarisation signals of exoplanets.

This paper is organised as follows. In Sec.~\ref{sec:sect_2}, we
describe our numerical method to calculate the flux and polarisation
of starlight that is reflected by a horizontally inhomogeneous planet,
and in Sec.~\ref{sec:sect_3}, we present simulations of flux and
polarisation for different types of horizontal inhomogeneities.  In
Sec.~\ref{sec:sect_4}, we present flux and polarisation signals of
horizontally inhomogeneous Earth-like planets and compare them to
signals for horizontally homogeneous planets.  Finally, in
Sec.~\ref{sec:sect_5}, we discuss and summarize our results.
Appendix~A contains the results of testing our code for horizontally
inhomogeneous planets against an existing code for horizontally
homogeneous planets.

%%%%%%%%%%%%%%%%%%%%%%%%%%%%%%%%%%%%%%%%%%%%%%%%%%%%%%%%%%%%%%%%%%%%%%%%%%
\section{Calculating reflected starlight}
\label{sec:sect_2}

%%%%%%%%%%%%%%%%%%%%%%%%%%%%%%%%%%%%%%%%%%%%%%%%%%%%%%%%%%%%%%%%%%%%%%%%%%

Light can fully be described by a flux vector $\pi\vec{F}$, as follows
%------------------------------------
\begin{equation}
   \pi\vec{F}= \pi \left[\begin{array}{c} 
               F \\ Q \\ U \\ V 
               \end{array}\right],
\label{eq:first}
\end{equation}
%------------------------------------
with $\pi F$ the total, $\pi Q$ and $\pi U$ the linearly and $\pi V$
the circularly polarised fluxes \citep[see e.g.][]{hansentravis74,
  hovenier04, stam08}. Parameters $\pi F$, $\pi Q$, $\pi U$ and $\pi
V$ depend on the wavelength $\lambda$, and have dimensions
W~m$^{-2}$m$^{-1}$. Parameters $\pi Q$ and $\pi U$ are defined with
respect to a reference plane, for which we choose the planetary
scattering plane, i.e. the plane through the centers of the planet,
star and observer. Note that this plane is usually not the same as the
planetary orbital plane; only for orbits that are seen edge--on, the
two planes coincide at all phase angles. In the following, we will
ignore $\pi V$, because it is usually very small
\citep[][]{hansentravis74}, and because the errors in calculated
values of $\pi F$, $\pi Q$, and $\pi U$ due to ignoring $\pi V$ are
negligible \citep{stam05}.

The degree of
polarisation $P$ of vector $\pi \vec{F}$ is defined as
%------------------------------------
\begin{equation}
   P= \frac{\sqrt{Q^{2}+U^{2}}}{F},
\label{eq:poldef}
\end{equation}
%------------------------------------
which is independent of the choice of reference plane.  For planets
that are mirror--symmetric with respect to the planetary scattering
plane (i.e. horizontally homogeneous planets), Stokes parameter $U$
equals zero. In that case, we can use an alternative definition of the
degree of polarisation
%------------------------------------
\begin{equation}
   P_{\rm s}= - \frac{Q}{F},
\label{eq:signedP}
\end{equation}
%------------------------------------
with the sign indicating the direction of the polarisation, i.e.  if
$P_{\rm s} > 0$ ($P_{\rm s} < 0$) the light is polarized perpendicular
(parallel) to the planetary scattering plane. The absolute value of
$P_{\rm s}$ is just equal to $P$.

We calculate the flux vector of starlight that has been reflected by a
spherical planet with radius $r$ at a distance $d$ from the observer
using ($d \gg r$) as \citep[see][]{stam06}
%------------------------------------
\begin{equation} 
   \pi \vec{F}(\alpha)=
   \frac{1}{4}\frac{r^2}{d^2}\vec{S}(\alpha) \pi \vec{F}_0,
\label{eq:fluxSF}
\end{equation}
%------------------------------------
with $\alpha$ the planetary phase angle, i.e.~the angle between the
star and the observer as seen from the planet's center. Furthermore,
$\pi \vec{F}_0$ is the flux vector of the incident starlight and
$\vec{S}$ the $4\times4$ planetary scattering matrix.  
In the following, we normalise Eq.~\ref{eq:fluxSF} assuming $r=1$ and
$d=1$.

As described in \citet{stam06}, the planetary scattering matrix
$\vec{S}$ can be calculated by integrating local reflection matrices
$\vec{R}$ across the illuminated and visible part of the planetary
disk, as follows
%------------------------------------
\begin{equation}
\label{eq:matS}
\vec{S}(\alpha)= \frac{4}{\pi} \int_{\rightmoon} \mu \hs \mu_0 \hs
                 \vec{L}(\beta_2) \hs
                 \vec{R}(\mu,\mu_0,\Delta \phi)
                 \vec{L}(\beta_1) \hs \mathrm{d}O,
\end{equation}
%------------------------------------
where $\mathrm{d}O$ is a surface element on the planet, and $\vec{R}$
is the local reflection matrix, which describes how starlight that is
incident on a given location of the planet is reflected towards the
observer. The reference planes for $\vec{R}$ are the local meridian
planes, which contain the direction of propagation of the incident and
reflected light, respectively, and the local vertical direction. The
matrices $\vec{L}$ are so--called rotation matrices
\citep[see][]{hovenier04,hovenier83} that are used to rotate from the
planetary scattering plane to the local meridian planes and back:
%------------------------------------
\begin{equation}
\label{eq:matL}
\vec{L}(\beta)=\left[\begin{array} {c c c c} 1 & 0 &0 &0 \\ 0&
    \cos2\beta &\sin2\beta & 0 \\ 0 & -\sin2\beta & \cos2\beta & 0
    \\ 0 & 0 & 0 & 1\end{array} \right],
\end{equation}
%------------------------------------
Angle $\beta$ is measured in the anti-clockwise direction from the old
to the new reference plane when looking in the direction of
propagation of the light.

In Eq.~\ref{eq:matS}, we assume that the planetary atmosphere and
surface (if present) are \textit{locally} plane-parallel and
rotationally symmetric with respect to the local vertical direction.
Therefore, each matrix $\vec{R}$ depends on $\mu_0= \cos \theta_0$,
with $\theta_0$ the angle between the local zenith and the direction
towards the star, on $\mu= \cos \theta$, with $\theta$ the angle
between the local zenith and the direction towards the observer, and
on $\Delta \phi = \phi-\phi_0$, the angle between the azimuthal angles
of the incident and the reflected light, respectively.  Azimuthal
angles are measured rotating clockwise when looking up from an
arbitrary, local vertical plane towards the local vertical plane
containing the direction of propagation of the light.

For \textit{horizontally homogeneous} planets, \citet{stam06} present
an efficient method for evaluating Eq.~\ref{eq:matS} that uses an
adding-doubling radiative transfer algorithm \citep[][]{dehaan87} to
calculate the coefficients of the expansion of the local reflection
matrix (which is the same across the planet) into a Fourier series.
These coefficients are then used to compute coefficients of the
expansion of matrix $\vec{S}$ into generalized spherical functions.
With these expansion coefficients, $\vec{S}$ can be calculated rapidly
for any phase angle $\alpha$. Since with this method, a planet is
basically treated as a single light--scattering particle, it cannot be
used for horizontally inhomogeneous planets.

To calculate $\vec{S}$ for \textit{horizontally inhomogeneous}
planets, we divide a planet into pixels small enough for the local
atmosphere and surface (if present) to be considered both
plane-parallel and horizontally homogeneous. For each type of pixel (a
combination of surface and atmosphere properties), we first calculate
the coefficients of the expansion of the local reflection matrix
$\vec{R}$ into a Fourier series, using the adding--doubling algorithm
\citep[][]{dehaan87}.  Then, for each given planetary phase angle and
each pixel, we use the respective Fourier coefficients to calculate
the local reflection matrices \citep[see][]{dehaan87}.  The local
matrices are summed up according to
%------------------------------------
\begin{equation} 
   \vec{S}(\alpha)= \frac{4}{\pi} \sum_{i=1}^{N} 
   \mu_i \hs \mu_{0i} \hs \vec{L}(\beta_{2i}) 
                          \vec{R}_i(\mu_i,\mu_{0i},\Delta \phi_i)
                          \vec{L}(\beta_{1i}) \mathrm{d}O_i,
\label{eq:scat_mat_code}
\end{equation}
%------------------------------------
with $N$ the total number of pixels on the illuminated and visible
part of the planetary disk.

In the following, we assume unpolarised incident starlight, since
integrated over the stellar disk, light of solar-type stars can be
assumed to be unpolarised \citep{kemp87}.  In this case, $\pi
\vec{F}_0 = \pi F_0 \vec{1}$, with $\vec{1}$ the unit column vector
and $\pi F_0$ the total incident stellar flux (measured perpendicular
to the propagation direction of the light), for which we assume a
normalized value of 1~W~m$^{-2}$~m$^{-1}$. Thanks to the assumption of
unpolarized incident starlight, rotation matrix $\vec{L}(\beta_{1i})$
can be ignored in Eq.~\ref{eq:scat_mat_code}.

Because of the normalizations and the assumption of unpolarized
incident light, the total flux that is reflected by a planet
(cf. Eq.~\ref{eq:fluxSF}) is given by
%------------------------------------
\begin{equation}
   \pi F_{\mathrm{n}}(\lambda,\alpha) = \frac{1}{4} a_{11}(\lambda,\alpha),
\label{eq:fn}
\end{equation}
%------------------------------------
with $a_{11}$ the (1,1)-element of matrix ${\bf S}$
\citep[see][]{stam08,karalidi11}. The subscript ${\rm n}$ indicates
the normalization. When $\alpha=0^\circ$, the hence normalized total
flux equals the planet's geometric albedo $A_{\mathrm{G}}$.  Our
normalized total and polarized fluxes $\pi F_{\mathrm{n}}$, $\pi
Q_{\rm n}$ and $\pi U_{\rm n}$ can straightforwardly be scaled to any
given planetary system using Eq.~\ref{eq:fluxSF} and inserting the
appropriate values for $r$, $d$ and $\pi F_0$.  The degree of
polarization ($P$ or $P_{\mathrm{s}}$) is independent of $r$, $d$ and
$\pi F_0$, and would thus not require any scaling.

We have tested our disk--integration code by comparing its results
with those of the code for horizontally homogeneous planets by
\citet{stam06} (see Appendix~\ref{appendix_A}).  We have not compared
it against other codes for modelling signals of horizontally
inhomogeneous planets \citep[see e.g.][]{ford01, oakley09,
  tinetti06b}, because these codes ignore polarisation, which is the
most interesting feature of our code. A comparison between calculated
total fluxes would require us to disable the polarisation calculations
in our adding-doubling code, since ignoring polarisation introduces
errors of up to several percent in total flux calculations
\citep[][]{stam05}.  From the comparison with the code for
horizontally homogeneous planets applied to horizontally inhomogeneous
planets using the weighted sum approximation, in which weighted sums
of flux vectors reflected by horizontally homogeneous planets are used
to approximate the flux vectors of horizontally inhomogeneous planets
(see Appendix~\ref{appendix_A}), we conclude that our code is accurate
enough for application to horizontally inhomogeneous planets, provided
enough pixels are used across the disk, not only for resolving the
spatial variations but also the variations in illumination and viewing
angles across pixels.

%%%%%%%%%%%%%%%%%%%%%%%%%%%%%%%%%%%%%%%%%%%%%%%%%%%%%%%%%%%%%%%%%%%%%%%%%%
\section{Sensitivity to horizontal inhomogeneities}
\label{sec:sect_3}

%%%%%%%%%%%%%%%%%%%%%%%%%%%%%%%%%%%%%%%%%%%%%%%%%%%%%%%%%%%%%%%%%%%%%%%%%%

In this section, we present flux and polarisation (degree and angle)
signals of planets with different types of horizontal inhomogeneities
on their surfaces as calculated using our code to show the signatures
of inhomogeneities and the differences with signatures of horizontally
homogeneous planets with similar surface albedos. Unless stated
otherwise, each planet has a gaseous, Rayleigh scattering atmosphere
with a total optical thickness of $\sim 0.1$ (no absorption), and a
flat surface.  For the surface albedo, we choose the extreme values of
0.0 (black) and 1.0 (white), since they give the largest contrast in
flux and polarisation.

Establishing which types of inhomogeneities could still be handled
with e.g.\ a weighted sum approximation and which would need a full
scale horizontally inhomogeneous approach with e.g. our code (from
hereon: the HI-code), is interesting for saving computing time when
possible. The disadvantage of our HI-code is namely the large amount
of computing time it requires as compared to the horizontally
homogeneous code of \citet{stam06} (from hereon: the HH-code).  For
example, the HI-code takes about 10$^5$ times more time than the
HH-code for calculating the flux vectors of a horizontally homogeneous
planet covered by a cloud layer with optical thickness 2, for phase
angles $\alpha$ from 0$^\circ$ to 180$^\circ$ in steps of 2$^\circ$,
with the planet having been divided into pixels of $2^\circ \times
2^\circ$, and for a single wavelength.

The difference in computation times would be irrelevant if the
computing time of the HI-code were negligible.  Unfortunately this is
not the case. This time depends strongly on the properties of the
model atmosphere, especially on the absorption and scattering optical
thickness of the gases and particles in the atmosphere and on the
angular variation of the single scattering properties of the
scattering particles. In particular adding polarization to the flux
calculations increases the time by at least an order of magnitude, and
of course, the timing increases almost linearly with the number of
wavelengths at which calculations are required (since these
calculations are independent of each other, they could easily be done
in parallel).  For the cloudy model planet calculations described in
the previous paragraph, the HH-code takes about a minute on an average
workstation.

The difference in computing time is not spent in the radiative
transfer calculations themselves, since the two codes use the same
adding-doubling radiative transfer algorithm \citep[based
  on][]{dehaan87} and were run with the same numerical accuracy.
Also, because both planets were horizontally homogeneous, the
radiative transfer calculation had to be performed only once for each
code. Indeed, in the HI-code, the additional computing time is mostly
spent in the integration of the flux vectors across the planetary
disk.  In particular, for $\alpha=0^\circ$, the number of $2^\circ
\times 2^\circ$ pixels across the disk is more than 8000. For each
pixel, the appropriate Fourier coefficients have to be determined from
the list of calculated coefficients, they have to be summed to
calculate the local reflection matrix, and with that the locally
reflected flux vectors. Then, the locally reflected flux vectors have
to be rotated to the planetary scattering plane in order to be summed
up to calculate the reflected flux vector of the planet. Even though
the computing time per pixel can be relatively small, the mere number
of pixels (which of course depends on the planetary phase angle) can
result in long computing times.

%-------------------------------------------------------------------------
% Figure 1: made with MX_PL_PARAL/N_RES_PAP2/DATA/PZL/compare.pro
%-------------------------------------------------------------------------
\begin{figure}
\centering
\includegraphics[width=85mm]{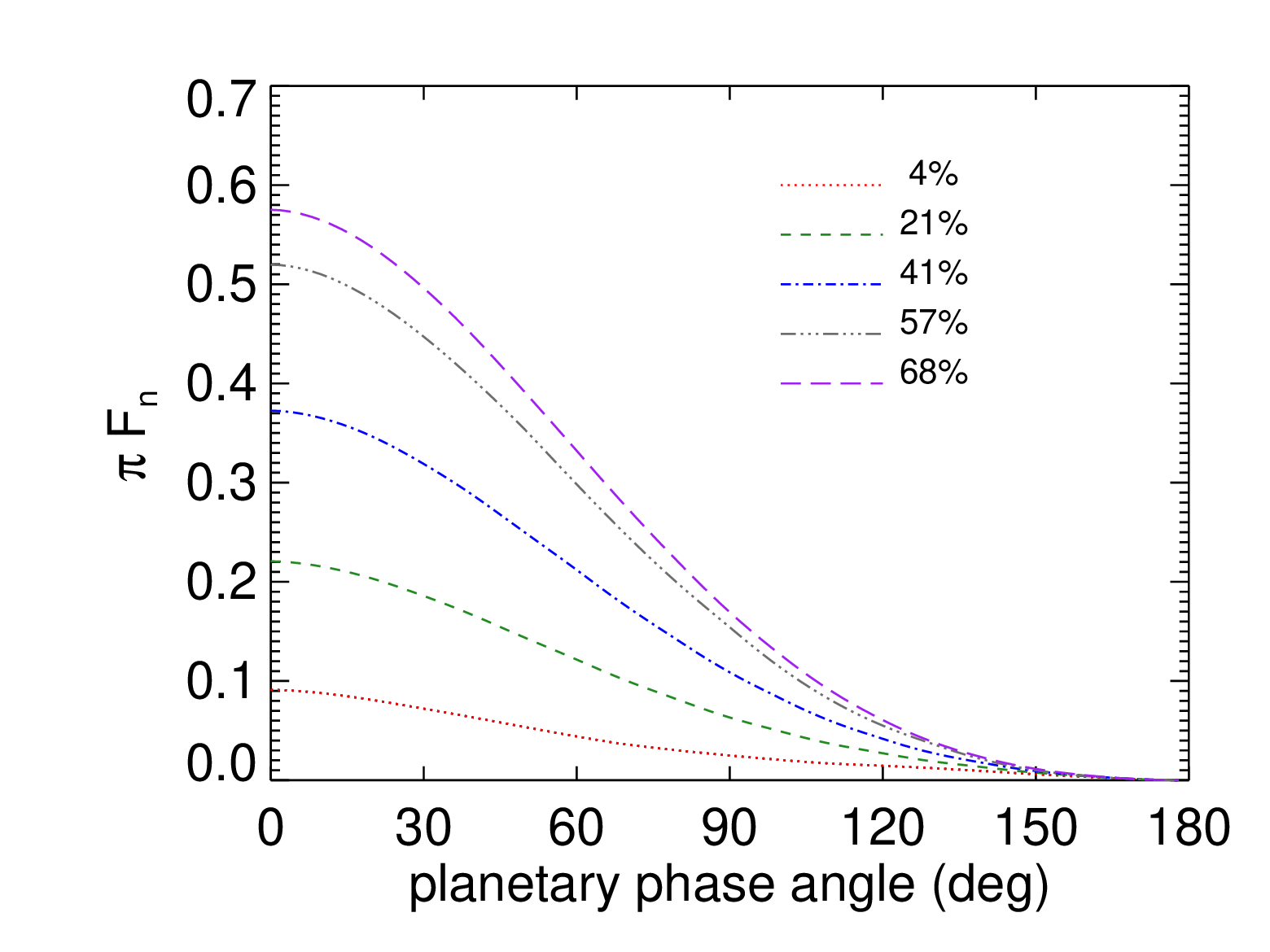}%{karalidi4a.ps}
\hspace{0.8cm}
\centering
\includegraphics[width=85mm]{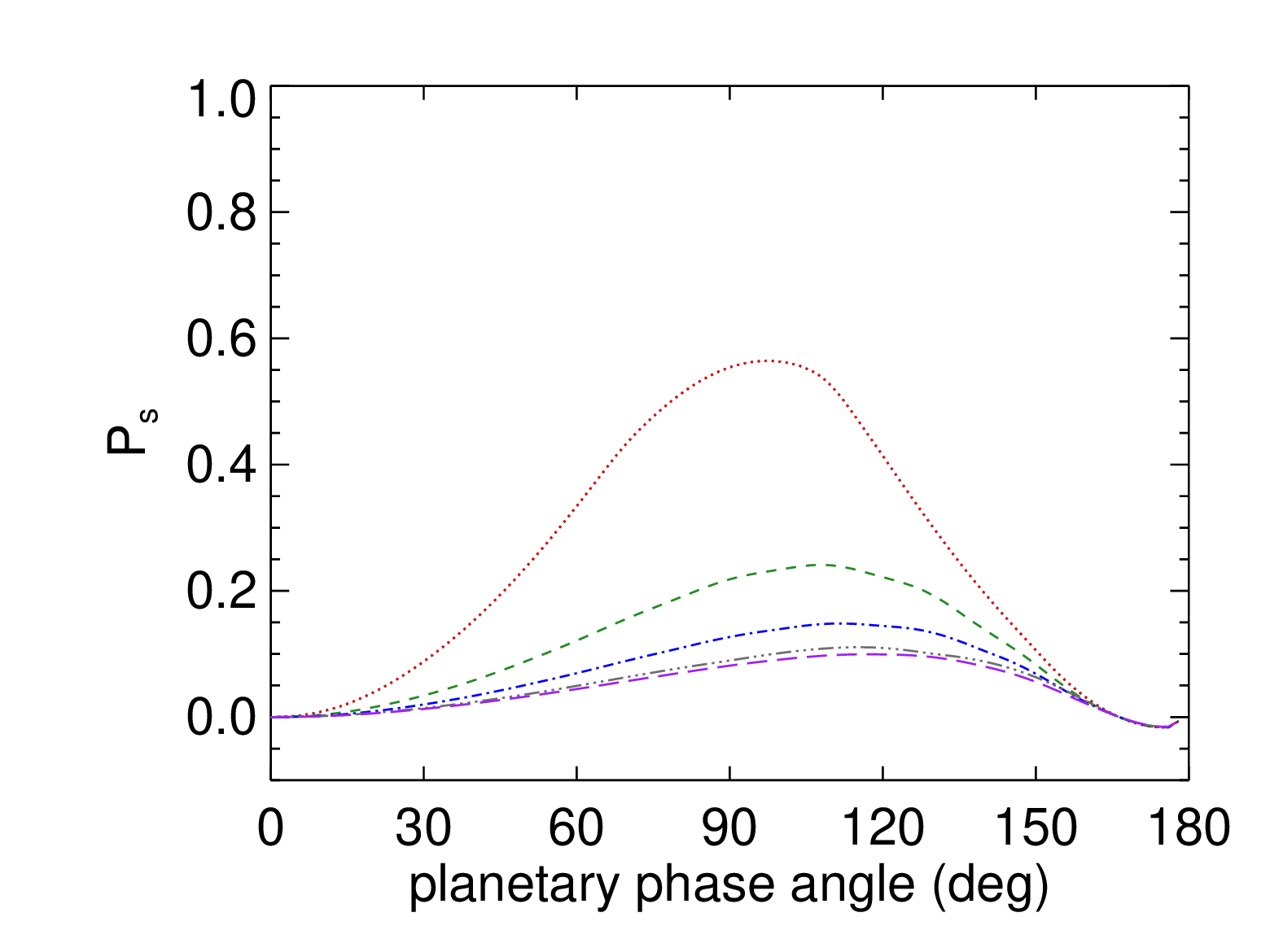}%{karalidi4b.ps}
\caption{$\pi F_\mathrm{n}$ and $P_\mathrm{s}$ as functions of
  $\alpha$ for black, cloud--free planets with white spots occupying
  4\% (red, dotted line), 21\% (green, dashed line), 41\% (blue,
  dashed--dotted line), 57\% (grey, dashed--triple--dotted line) and
  68\% (purple, long--dashed line) of the planet.}
\label{fig:pzl_01_fluxpol}
\end{figure}

%%%%%%%%%%%%%%%%%%%%%%%%%%%%%%%%%%%%%%%%%%%%%%%%%%%%%%%%%%%%%%%%%%%%%%%%%%
\subsection{Planets with spots}

The first type of horizontally inhomogeneous planets have black
surfaces with homogeneously distributed white spots. We start with
spots consisting of single white pixels that cover 4$\%$ of the
planet's surface, and let the area of each spot grow until the spots
cover 68$\%$ of the surface.  Figure~\ref{fig:pzl_01_fluxpol} shows
$\pi F_\mathrm{n}$ and $P_\mathrm{s}$ (see Eq.~\ref{eq:signedP}) as
calculated using the HI-code as functions of the planetary phase angle
$\alpha$ for these planets. As can be seen, increasing the percentage
of white pixels from 4$\%$ to 68$\%$, increases $\pi F_\mathrm{n}$
smoothly from 0.09 to 0.58 at $\alpha=0^\circ$, while $P_\mathrm{s}$
decreases from 0.6 to 0.08 at $\alpha=90^\circ$ (this phase angle does
not coincide with the maximum of $P_{\rm s}$). With increasing surface
brightness, the peak value of $P_\mathrm{s}$ shifts from
$\alpha=98^\circ$ (4$\%$ white) to $\alpha=116^\circ$ (68$\%$ white).

%-------------------------------------------------------------------------
% Figure 2: made with MX_PL_PARAL/N_RES_PAP2/DATA/PZL/compare.pro
%-------------------------------------------------------------------------
\begin{figure}
\centering
\includegraphics[width=85mm]{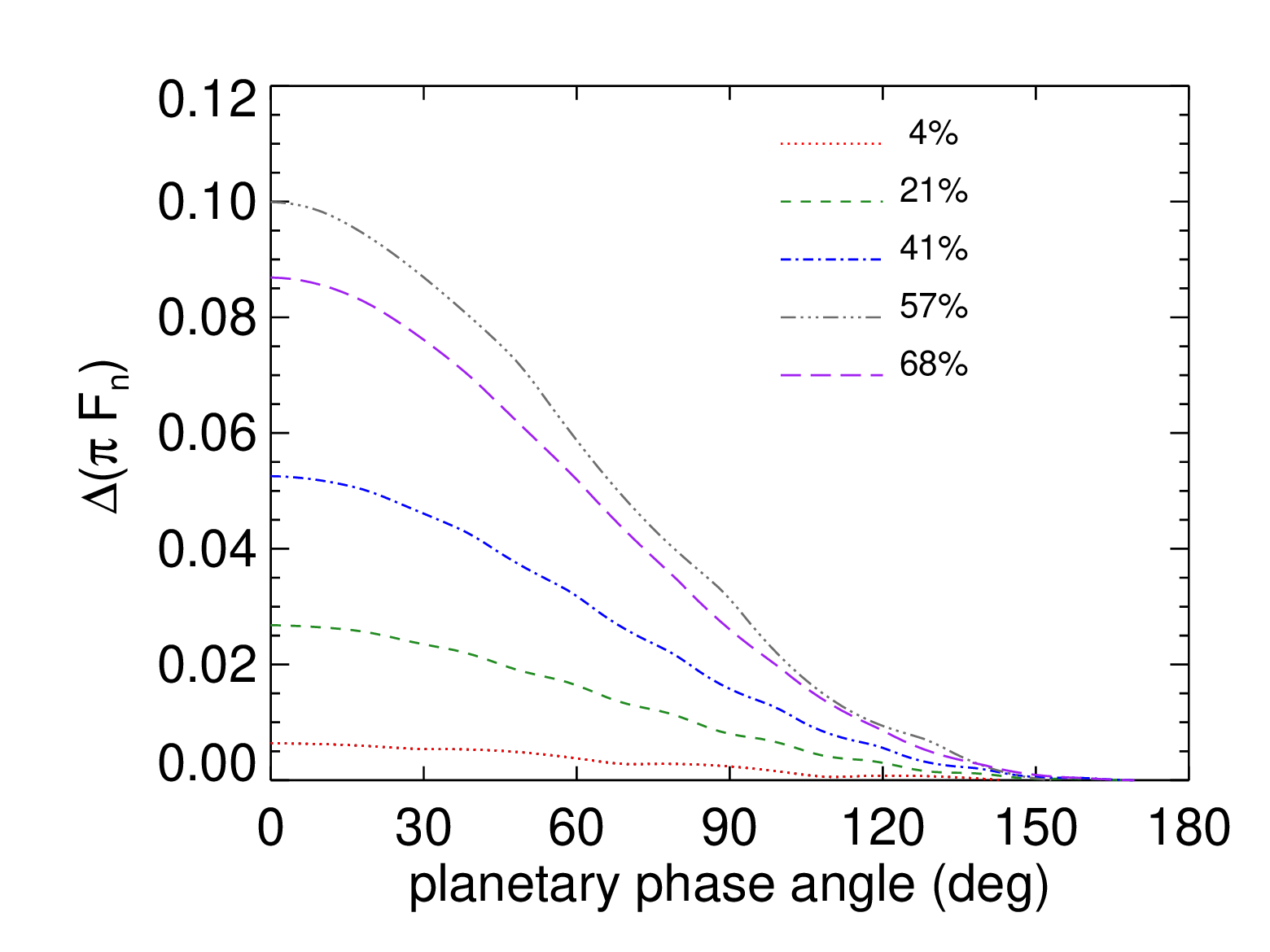}%{karalidi5a.ps}
\hspace{0.8cm}
\centering
\includegraphics[width=85mm]{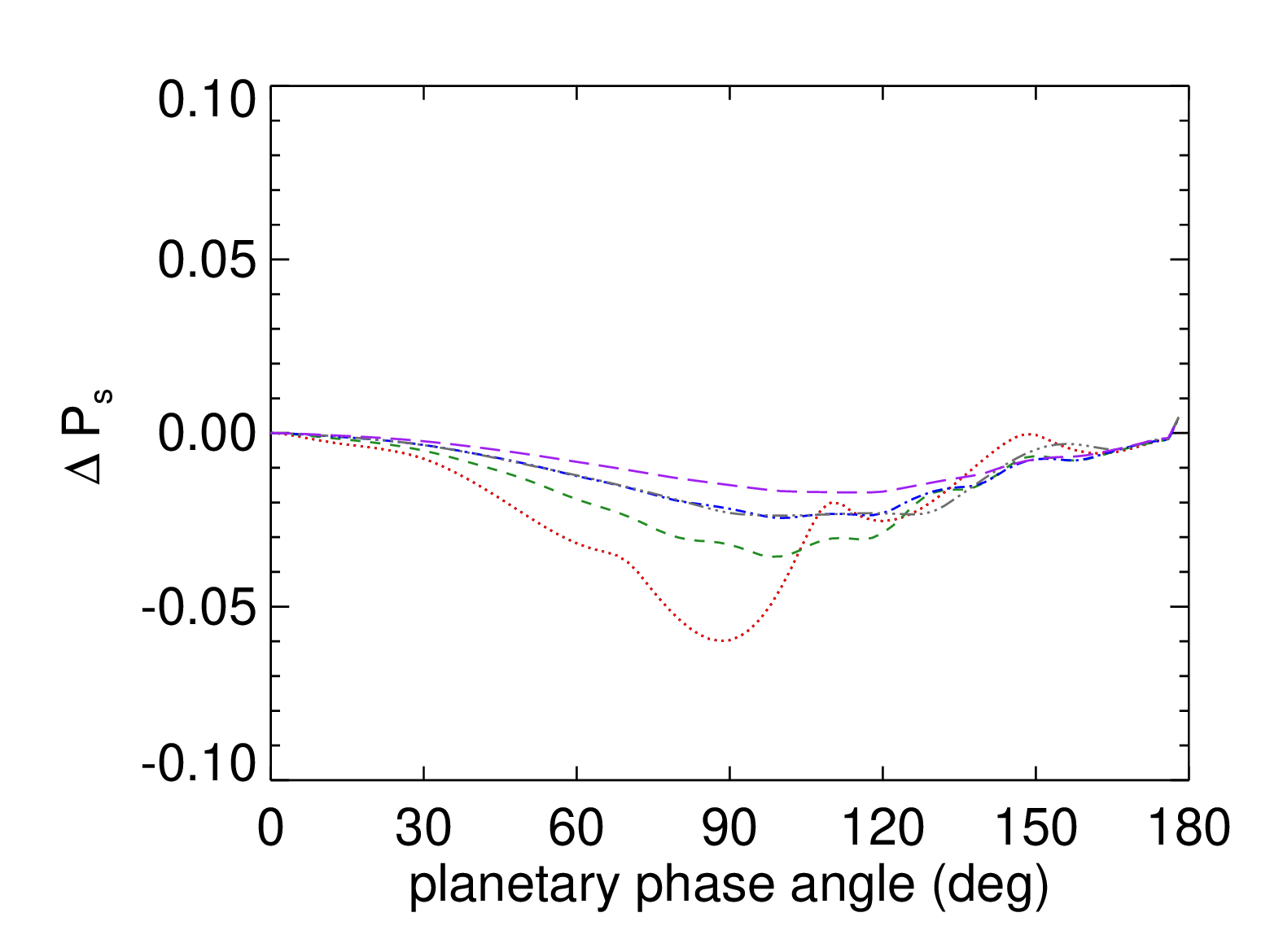}%{karalidi5b.ps}
\caption{Relative difference $\Delta \pi F_\mathrm{n}$ and the
  absolute difference $\Delta P_\mathrm{s}$ between the flux and
  polarisation phase functions of the white spotted planets of
  Fig.~\ref{fig:pzl_01_fluxpol} as calculated using the HI-code and
  the HH-code. The coverage of white spots is 4\% (red, dotted line),
  21\% (green, dashed line), 41\% (blue, dashed--dotted line), 57\%
  (grey, dashed--tripple--dotted line) and 68\% (purple, long--dashed
  line).}
\label{fig:pzl_01_fluxpol_var}
\end{figure}

Figure~\ref{fig:pzl_01_fluxpol_var} shows the difference between the
results from the HI-code and those from the HH-code (the latter
combined with with the weighted sum approximation, as described in
Appendix~\ref{appendix_A}).  In Fig.~\ref{fig:pzl_01_fluxpol_var}, it
is clear that when the planet is almost horizontally homogeneous (only
4$\%$ covered by white pixels), the maximum relative difference in
flux, $\Delta \pi F_\mathrm{n}$ is very small ($\sim0.64\%$ for
$\alpha=0^\circ$), with the HI-code giving the slightly higher
values. With increasing coverage, $\Delta \pi F_\mathrm{n}$ increases,
showing a maximum value of 10$\%$ for 57$\%$ coverage in
Fig.~\ref{fig:pzl_01_fluxpol_var}, to decrease again when the planet
is almost homogeneously white.

Because the degree of polarisation is itself a relative measure, we
show the differences in $P_\mathrm{s}$ between the HI-code and the
HH-code as absolute differences. As can be seen in
Fig.~\ref{fig:pzl_01_fluxpol_var}, the absolute difference $\Delta
P_\mathrm{s}$ is largest ($\sim -6\%$ percent point around 90$^\circ$)
for the darkest planet and decreases towards the whiter planets. The
HH-code produces at (almost) all phase angles a larger $P_{\rm s}$
than the HI-code. Interestingly, the polarisation phase curve of the
darkest planet is significantly less symmetrical when horizontally
inhomogeneities are accurately taken into account (HI-code) than with
the HH-code and the weighted sum approximation.

%-------------------------------------------------------------------------
\subsection{Planets with center continents}

\begin{figure}
\centering
\includegraphics[width=85mm]{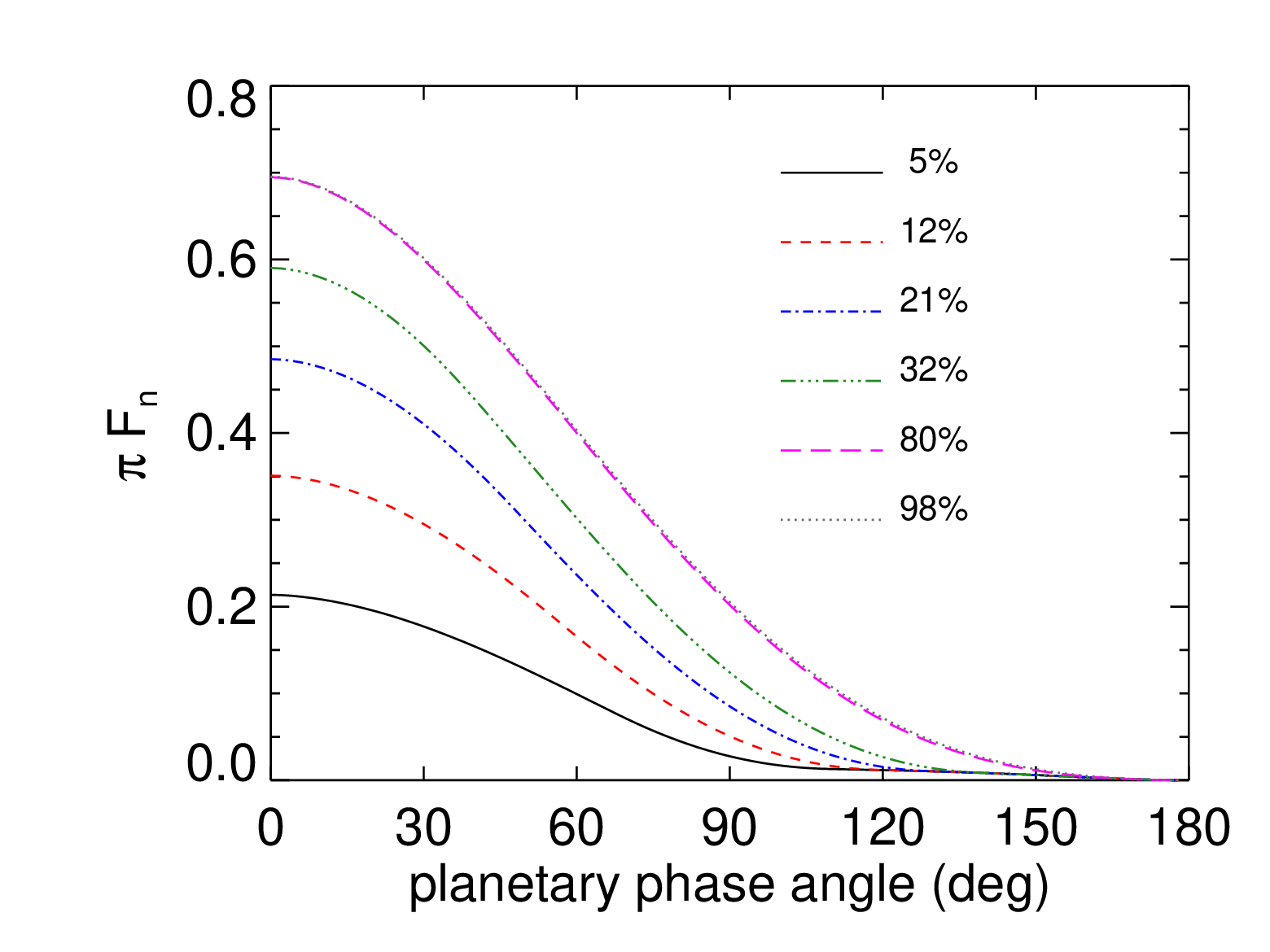}%{karalidi6a.ps}
\hspace{0.8cm}
\centering
\includegraphics[width=85mm]{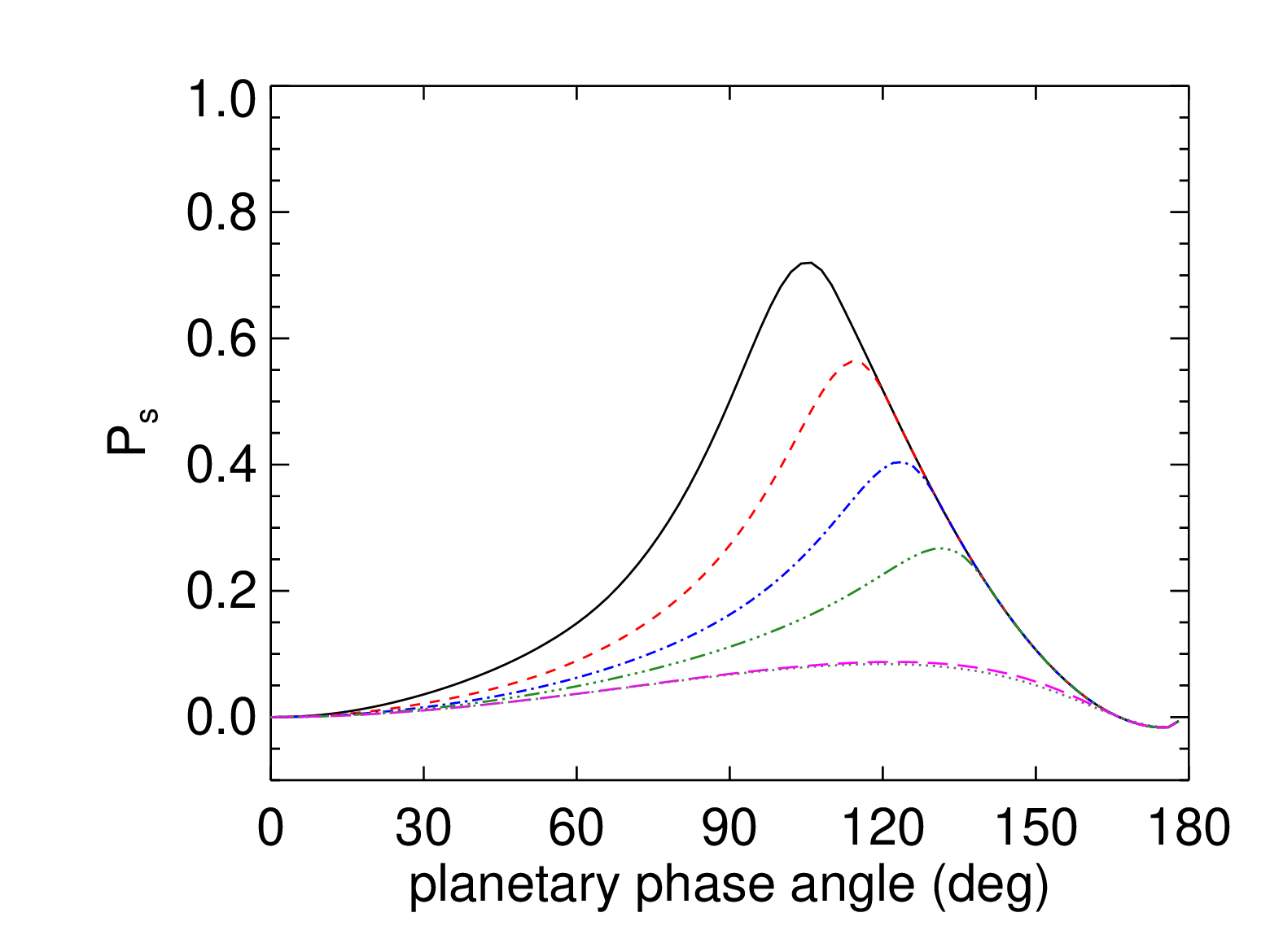}%{karalidi6b.ps}
\caption{$\pi F_\mathrm{n}$ and $P_\mathrm{s}$ as functions of
  $\alpha$ for black, cloud--free planets with a white continent on
  the center of the planetary disk facing the observer, for different
  coverages of the continent: 5\% (black, solid line), 12\% (red,
  dashed line), 21\% (blue, dashed--dotted line), 32\% (green,
  dashed--triple--dotted line), 80\% (magenta, long--dashed line) and
  98\% (gray, dotted line).}
\label{fig:spot_01_fluxpol}
\end{figure}

The second type of horizontally inhomogeneous planets have black
surfaces and a circular ``continent'' of white pixels at the center of
the planetary disk facing the observer (rotation of the planet, hence
the rotation of the continent in and out of the field-of-view is not
taken into account here).  Figure~\ref{fig:spot_01_fluxpol} shows $\pi
F_\mathrm{n}$ and $P_\mathrm{s}$ as functions of $\alpha$ for planets
with continents that cover from 5\% to 98\% of the disk.  Both for the
flux and the polarisation, the curves for 80\% and 98\% coverage
virtually overlap, because in these cases, the black pixels are all
located along the limb of the planet and hardly contribute to the
total signal.  The flux phase functions in
Fig.~\ref{fig:spot_01_fluxpol} have very similar shapes as those in
Fig.~\ref{fig:pzl_01_fluxpol}, although the latter are darker for the
same surface coverage of white pixels, which is not surprising since
they have more black pixels on the front side of the disk.  The
polarisation phase curves in Fig.~\ref{fig:spot_01_fluxpol} clearly
have different, more asymmetrical shapes than those in
Fig.~\ref{fig:pzl_01_fluxpol}, except for the largest coverages.  The
peak in the polarisation phase curves indicates the phase angle where
the continent disappears into the planet's nightside.

%-------------------------------------------------------------------------
% Figure 3: made with MX_PL_PARAL/N_RES_PAP2/DATA/SPOT/compare.pro
%-------------------------------------------------------------------------

\begin{figure}
\centering
\includegraphics[width=85mm]{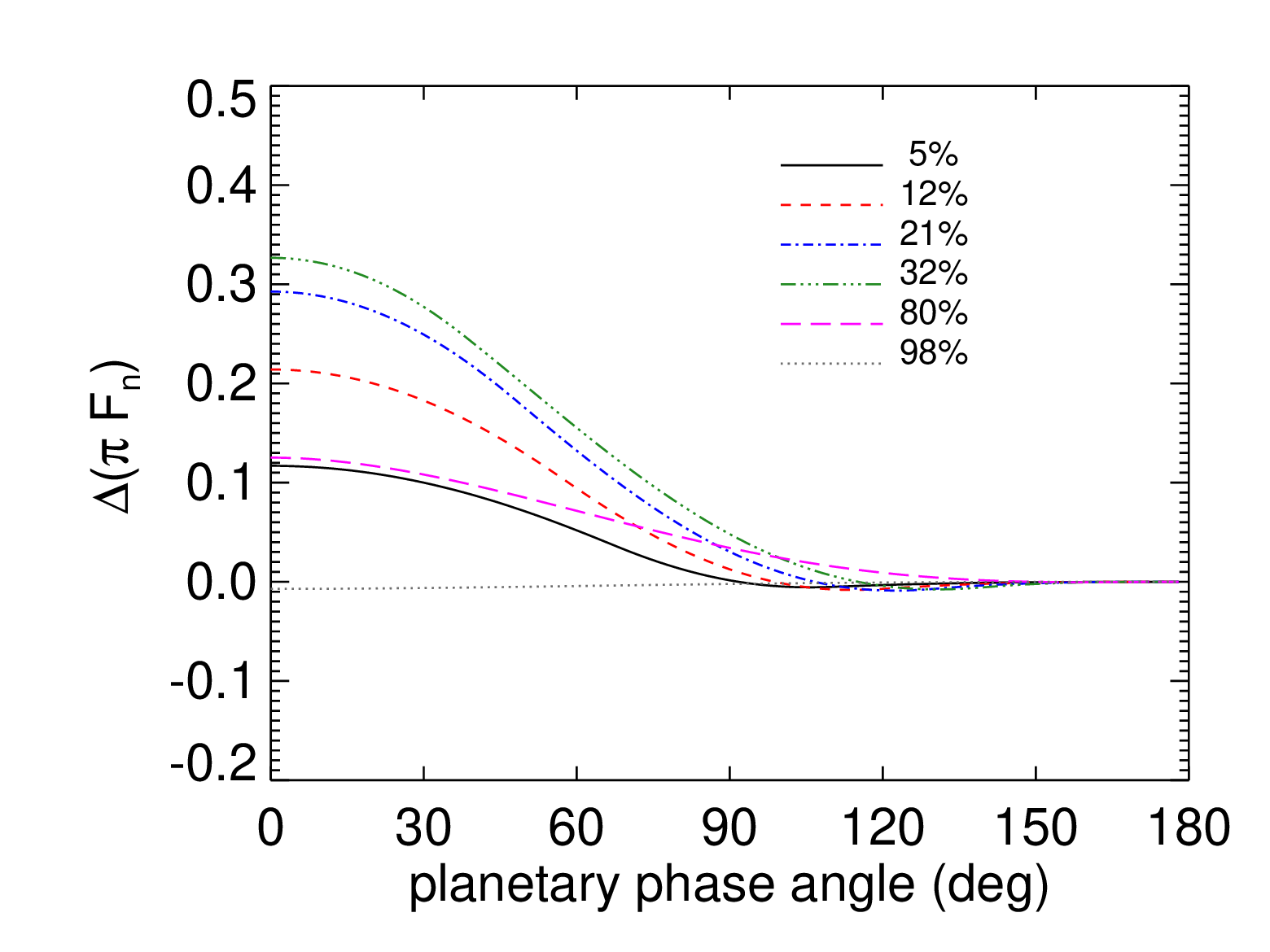}%{karalidi7a.ps}
\hspace{0.8cm}
\centering
\includegraphics[width=85mm]{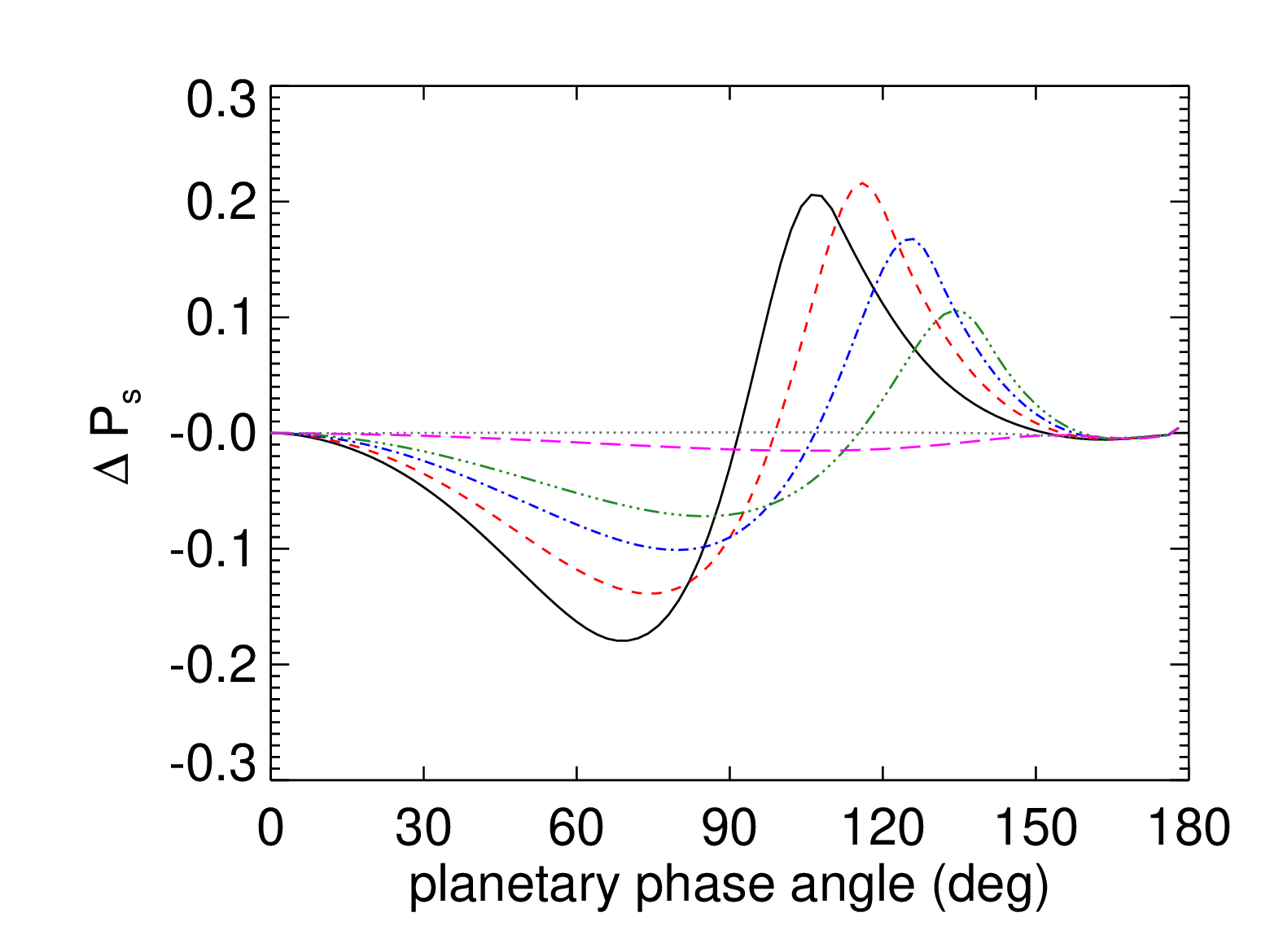}%{karalidi7b.ps}
\caption{$\Delta \pi F_\mathrm{n}$ and $\Delta P_\mathrm{s}$ between
  the phase functions of the white continent planets of
  Fig.~\ref{fig:spot_01_fluxpol} as calculated using the HI-code and
  the HH-code.  The coverage of the continents is: 5\% (black, solid
  line), 12\% (red, dashed line), 21\% (blue, dashed--dotted line),
  32\% (green, dashed--triple--dotted line), 80\% (magenta,
  long--dashed line) and 98\% (gray, dotted line) of the planetary
  disk.}
\label{fig:spot_01_fluxpol_var}
\end{figure}

Comparing the flux and polarisation signals of the planets with white
continents as calculated with our HI-code to the signals calculated
using the HH-code (Fig.~\ref{fig:spot_01_fluxpol_var}), it is obvious
that the differences $\Delta \pi F_\mathrm{n}$ and $\Delta
P_\mathrm{s}$ are smallest when the planetary disk is almost
completely covered by the continent.  In
Fig.~\ref{fig:spot_01_fluxpol_var}, the largest difference in the flux
is about 33$\%$, for a coverage of 32$\%$ and at $\alpha=0^\circ$,
with the HI-code giving the higher fluxes.  The difference in
polarisation, $\Delta P_{\rm s}$, clearly shows the strong asymmetry
around $\alpha=90^\circ$ of $P_\mathrm{s}$ when calculated with the
HI-code.

%-------------------------------------------------------------------------
% Figure 4: made with MX_PL_PARAL/N_RES_PAP2/DATA/SPOT/compare.pro
%-------------------------------------------------------------------------

\begin{figure}
\centering
\includegraphics[width=85mm]{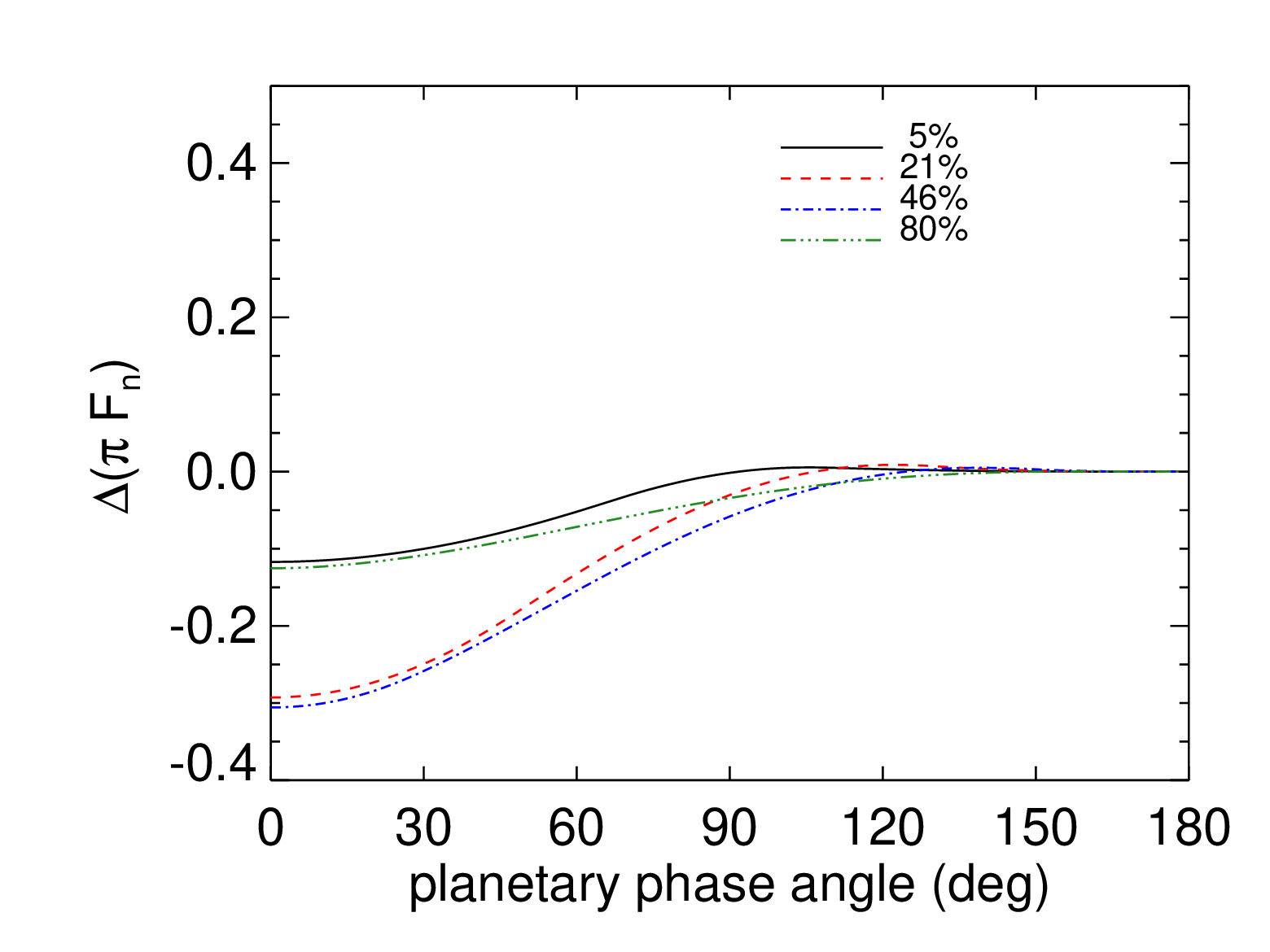}%{karalidi8a.ps}
\hspace{0.8cm}
\centering
\includegraphics[width=85mm]{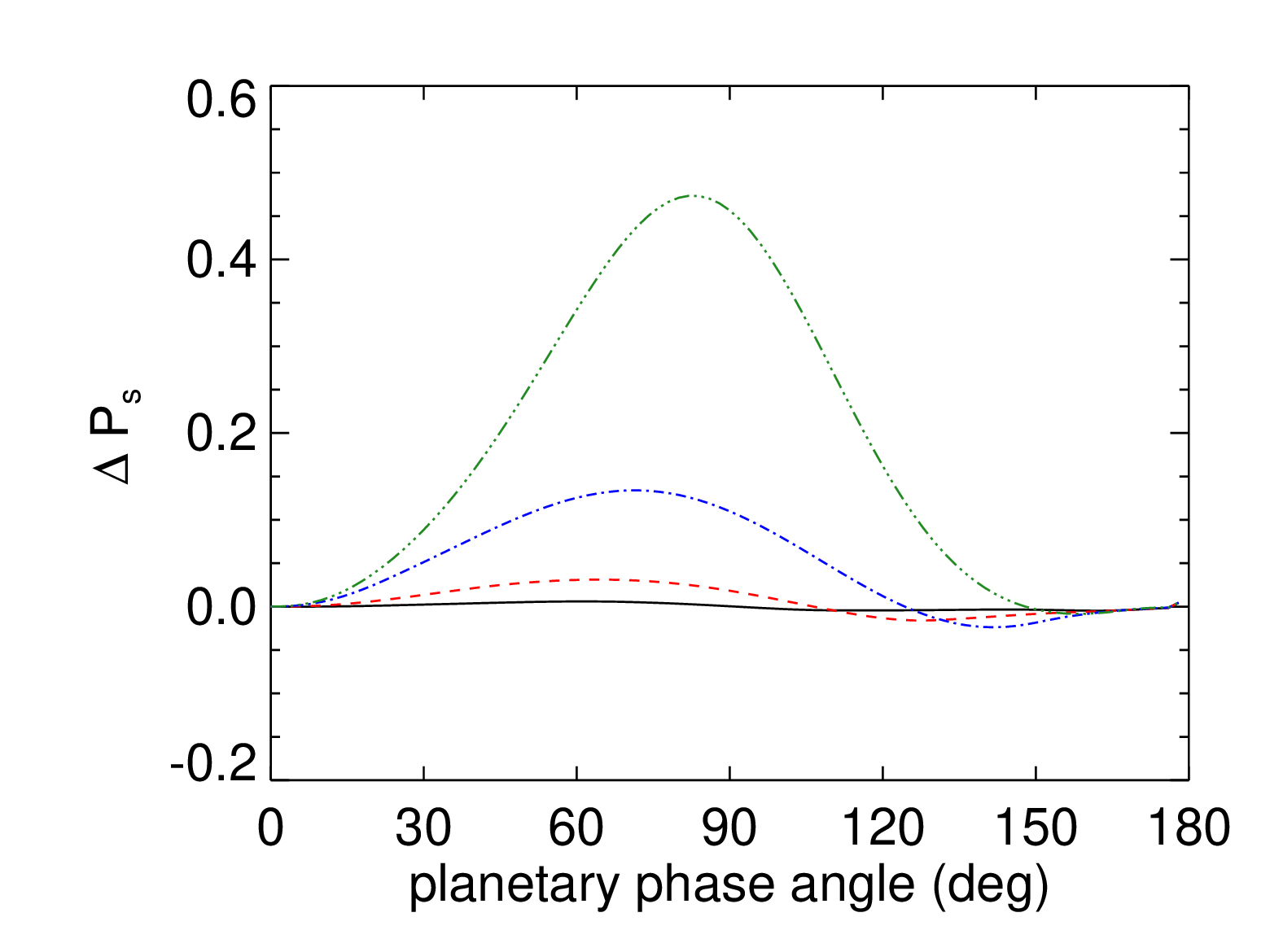}%{karalidi8b.ps}
\caption{Similar to Fig.~\ref{fig:spot_01_fluxpol_var}, except for
  white, cloud-free planets with black continents.  The coverage of
  the continents is: 5\% (black, solid line), 21\% (red, dashed line),
  46\% (blue, dashed--dotted line), and 80\% (green,
  dashed--triple--dotted line) of the planetary disk.}
\label{fig:negspot_01_fluxpol_var}
\end{figure}

Figure~\ref{fig:negspot_01_fluxpol_var} shows $\Delta \pi
F_\mathrm{n}$ and $\Delta P_\mathrm{s}$ for the same planets except
with a white surface and a black continent.  Not surprisingly, for
these planets, at most phase angles, $\pi F_\mathrm{n}$ is smaller
when calculated with the HI-code than with the HH-code for each
percentage of coverage at most phase angles, because of the
concentration of black pixels in the centre of the disk.  The
difference $\Delta P_\mathrm{s}$ is less asymmetric than for the black
planets with white continents
(cf. Fig.~\ref{fig:spot_01_fluxpol}). The polarisation phase functions
as calculated using the two codes are thus similarly shaped.  The
maximum of the polarisation phase function, however, does depend
strongly on the code, and is thus sensitive to the distribution of
pixels across the disk, with the HI-code yielding much higher maximum
values of $P_\mathrm{s}$ than the HH-code, except for the smallest
coverages. Indeed, for a white planet with a black continent covering
80\% of the disk, $P_\mathrm{s}$ is almost 50\% higher (around
$\alpha=80^\circ$) calculated with the HI-code than with the HH-code.
Note that for this coverage, the difference in flux is relatively
small ($\sim -10$\%) and similar to that for a coverage of 5\%.  The
difference in sensitivity to the spatial distribution of albedo across
the planetary disk between flux and polarisation clearly illustrates
the strengths of combined flux and polarisation measurements.

%-------------------------------------------------------------------------
% Figure 5: made with MX_PL_PARAL/N_RES_PAP2/DATA/NEGSPOT/compare.pro
%-------------------------------------------------------------------------

%-------------------------------------------------------------------------
\subsection{Planets with hemispherical caps}

The third type of horizontally inhomogeneous planets have black
surfaces and two, equally sized white caps on opposing sides of the
planet.  We will present results for three locations of the caps:
0$^\circ$ (the caps cover the ``north'' and the ``south'' poles of the
planet), 45$^\circ$, and 90$^\circ$ (the caps are on the ``eastern''
and ``western'' sides of the planetary disk).  Planets with caps at
45$^\circ$ are not mirror--symmetric with respect to the planetary
scattering plane. Therefore, Stokes parameter $U$ will usually not be
zero. In the following, we will therefore use Eq.~\ref{eq:poldef} to
define the degree of linear polarisation instead of
Eq.~\ref{eq:signedP}.

%-------------------------------------------------------------------------
% Figure 6: made with MX_PL_PARAL/N_RES_PAP2/DATA/POLAR/compare.pro
%-------------------------------------------------------------------------
\begin{figure}
\centering
\includegraphics[width=85mm]{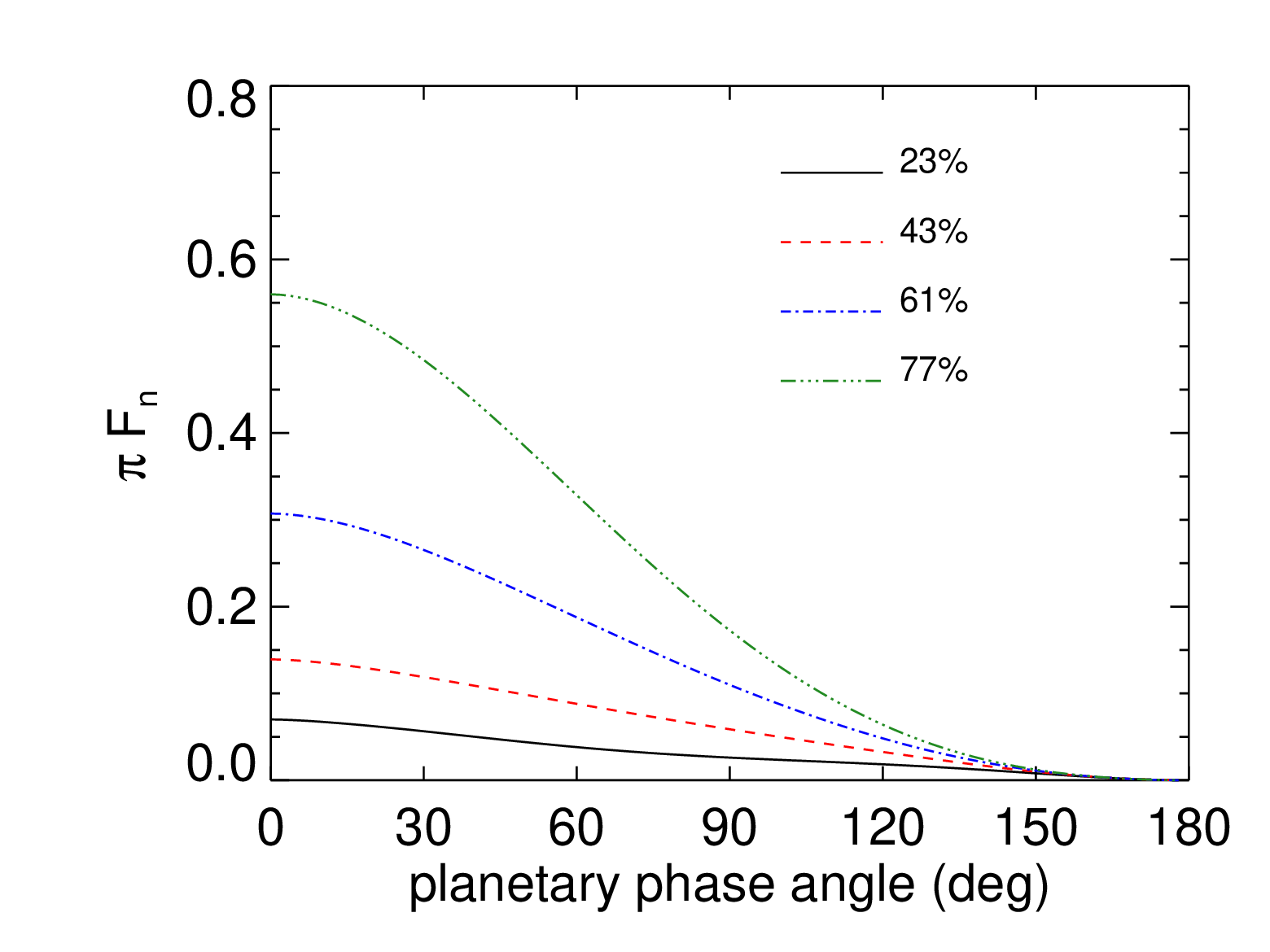}%{karalidi9a.ps}
\hspace{0.8cm}
\centering
\includegraphics[width=85mm]{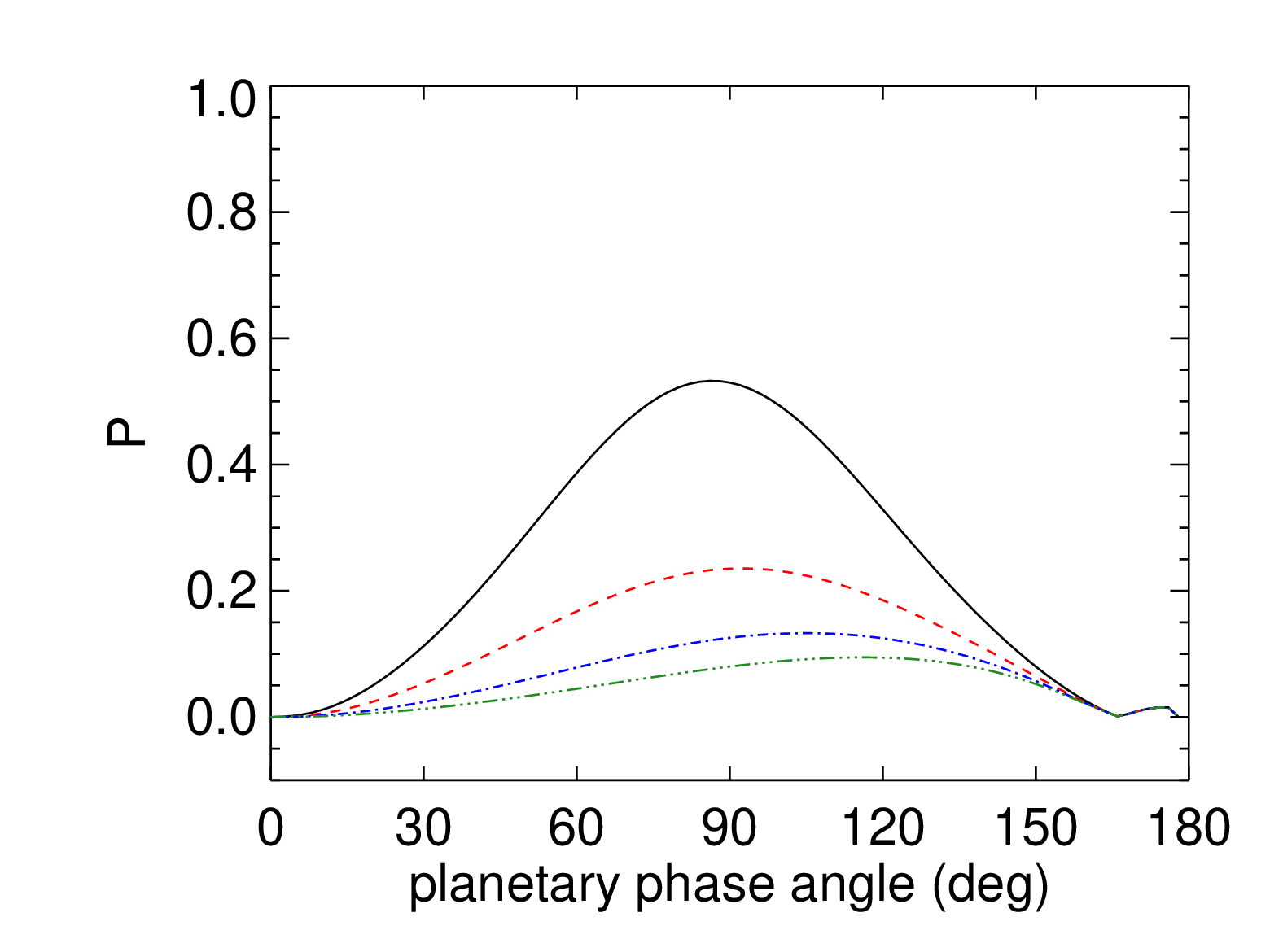}%{karalidi9b.ps}
\caption{$\pi F_\mathrm{n}$ and $P$ as functions of $\alpha$ for
  black, cloud--free planets with hemispherical caps occupying 23\%
  (black, solid line), 43\% (red, dashed line), 61\% (blue,
  dashed--dotted line), and 77\% (green, dashed--triple-dotted line)
  of the planet. The caps are located at the 'north' and 'south' poles
  of the planet.}
\label{fig:polcnt_01_fluxpol}
\end{figure}

First, we'll discuss the signals of planets with the caps covering the
poles of the planets. Figure~\ref{fig:polcnt_01_fluxpol} shows $\pi
F_\mathrm{n}$ and $P$ for different coverages of the caps as functions
of $\alpha$ as calculated using the HI-code.  The shape of the flux
phase function is very smooth, and at $\alpha=0^\circ$, $\pi
F_\mathrm{n}$ increases from 0.07 for 23\% coverage, to 0.56 for 77\%
coverage. In the latter case, the planet is basically white with a
black, equatorial belt (for a completely white planet, $\pi
F_\mathrm{n}$ would equal 0.7, see Fig.~\ref{fig:var_co_b_frst}).  The
polarisation phase function is quite symmetric for small caps, and $P$
decreases from almost 0.55 for 23\% coverage to 0.08 for 77\% coverage
at $\alpha=90^\circ$ (note that the maxima of the polarisation phase
functions occur at somewhat smaller and larger phase angles,
respectively).

%-------------------------------------------------------------------------
% Figure 7: made with MX_PL_PARAL/N_RES_PAP2/DATA/POLAR/compare.pro
%-------------------------------------------------------------------------
\begin{figure}
\centering
\includegraphics[width=85mm]{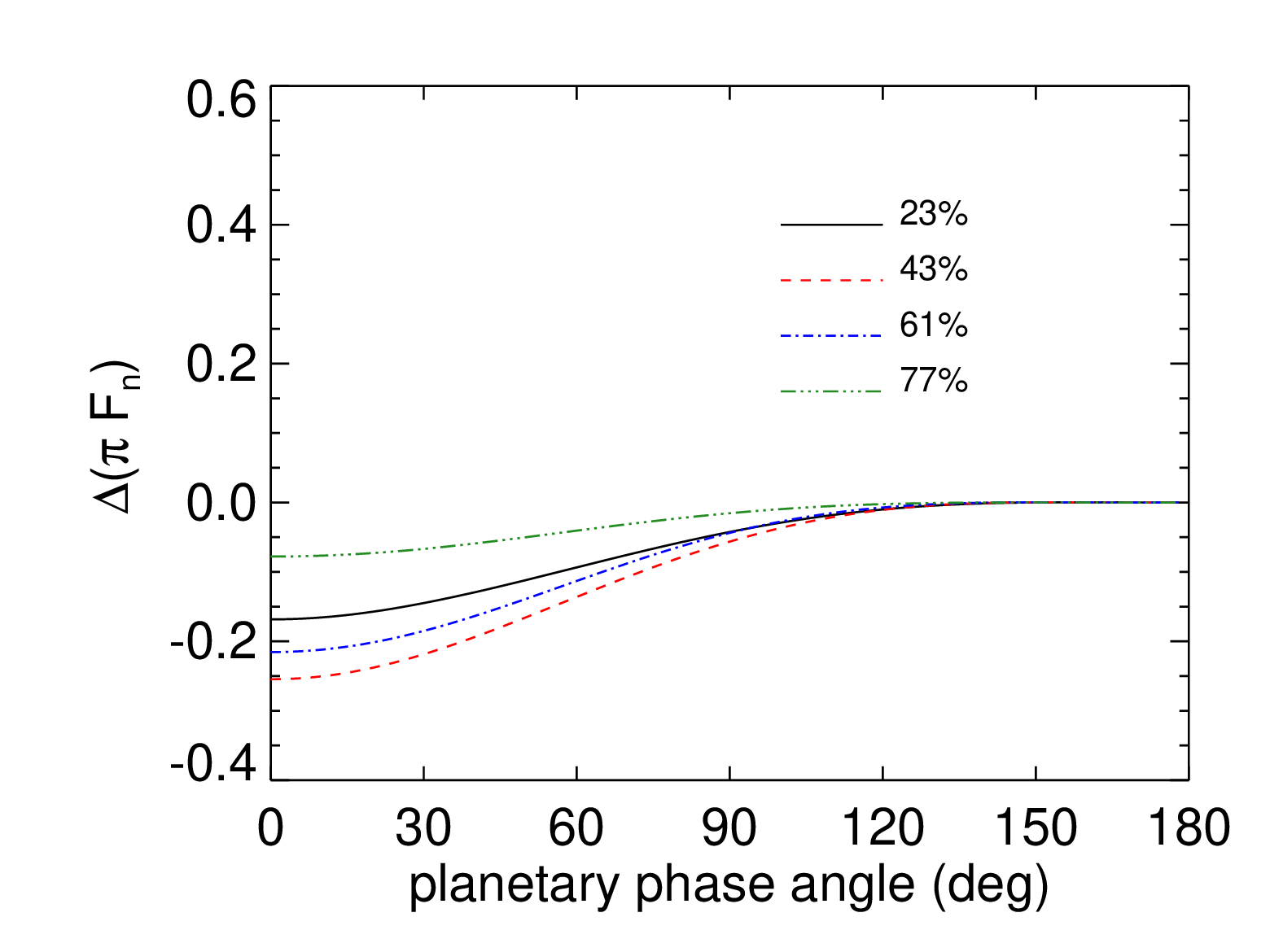}%{karalidi10a.ps}
\hspace{0.8cm}
\centering
\includegraphics[width=85mm]{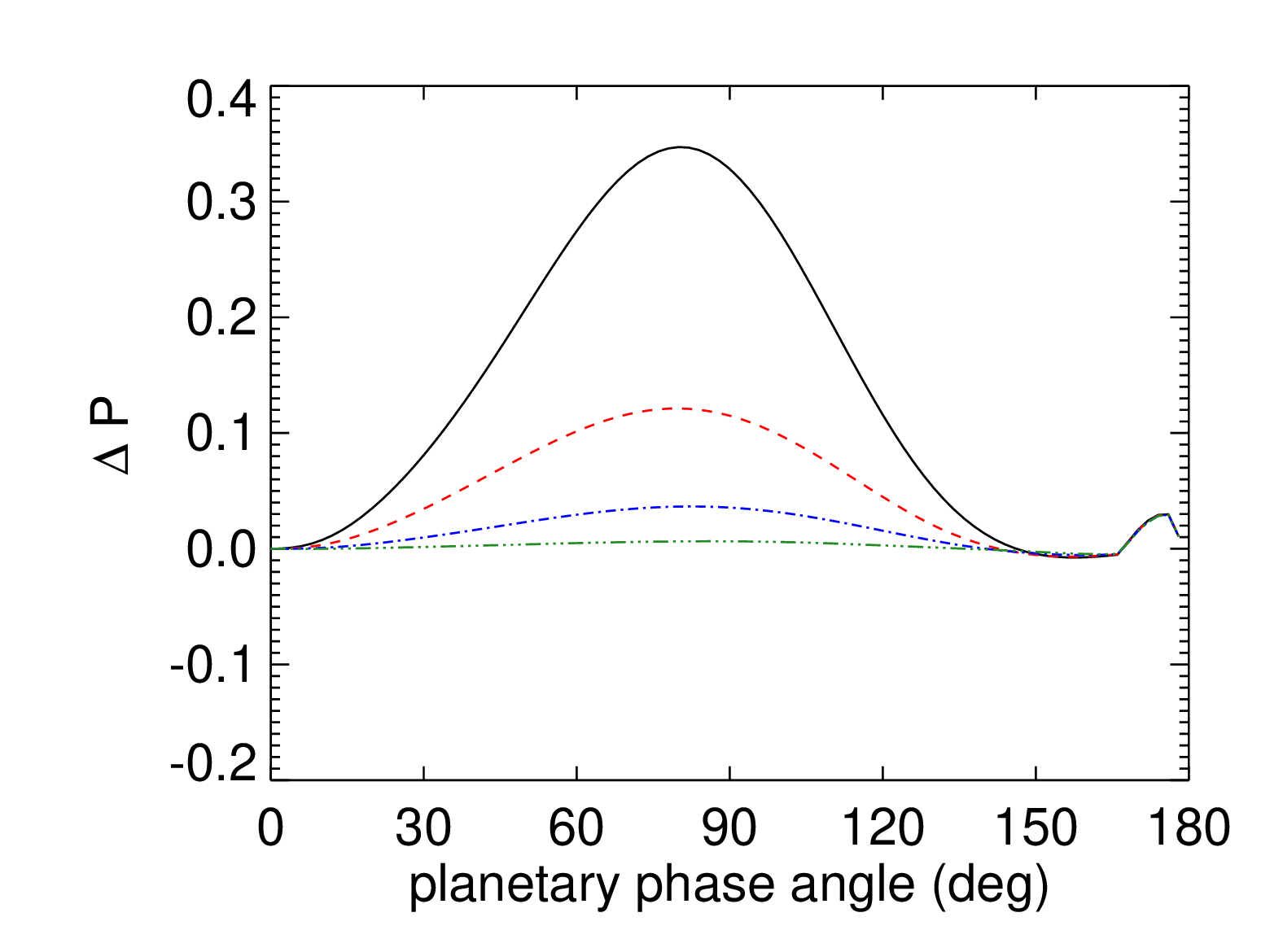}%{karalidi10b.ps}
\caption{$\Delta \pi F_\mathrm{n}$ and $\Delta P_\mathrm{s}$ between
  the phase functions of the capped planets of
  Fig.~\ref{fig:polcnt_01_fluxpol} as calculated using the HI-code and
  the HH-code.  The caps occupy 23\% (black, solid line), 43\% (red,
  dashed line), 61\% (blue, dashed--dotted line) and 77\% (green,
  dashed--triple-dotted line) of the planet.}
\label{fig:polcnt_01_fluxpol_var}
\end{figure}

In Fig.~\ref{fig:polcnt_01_fluxpol_var}, we show $\Delta \pi
F_\mathrm{n}$ and $\Delta P$ for the planets in
Fig.~\ref{fig:polcnt_01_fluxpol} when calculated using the HI-code and
the HH-code. The difference in flux is negative for all values of
$\alpha$: the HI-code thus gives smaller fluxes than the HH-code for
the same coverage. At $\alpha=0^\circ$, $\Delta \pi F_\mathrm{n}$ is
-0.14 for 23\% coverage, increases up to -0.21 for 43\% coverage, and
decreases again to -0.07 for 77\% coverage.  In differences in
polarisation show that for small coverages, the HI-code gives
significantly higher values of $P$ than the HH-code. The reason is of
course that for small coverages, the white pixels are mostly located
near the limb of the planetary disk, where they contribute little to
the total signal.
%-------------------------------------------------------------------------
% Figure 8: made with MX_PL_PARAL/N_RES_PAP2/DATA/POLAR/compare.pro
%-------------------------------------------------------------------------
\begin{figure}
\centering
\includegraphics[width=85mm]{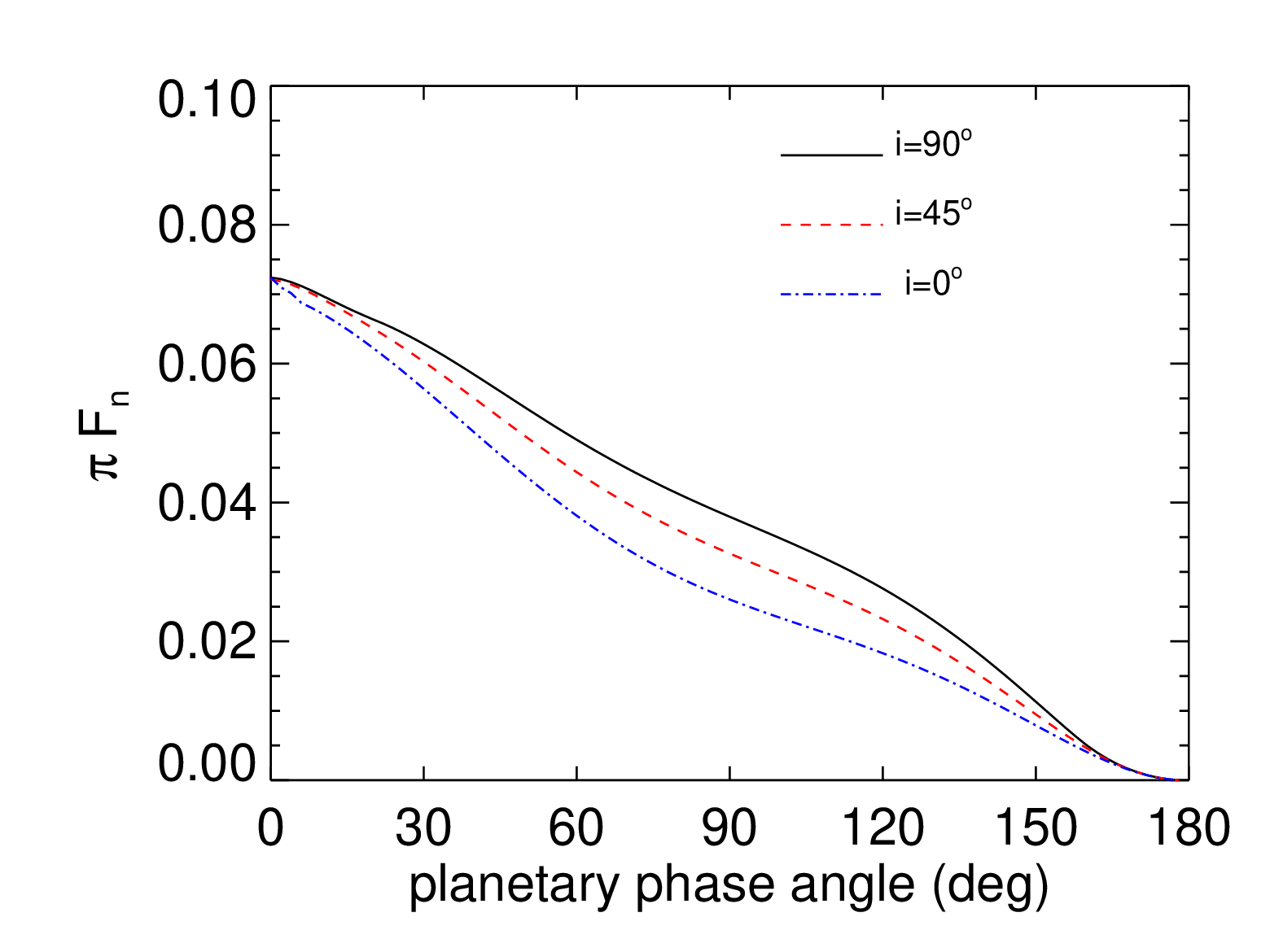}%{karalidi11a.ps}
\hspace{0.8cm}
\centering
\includegraphics[width=85mm]{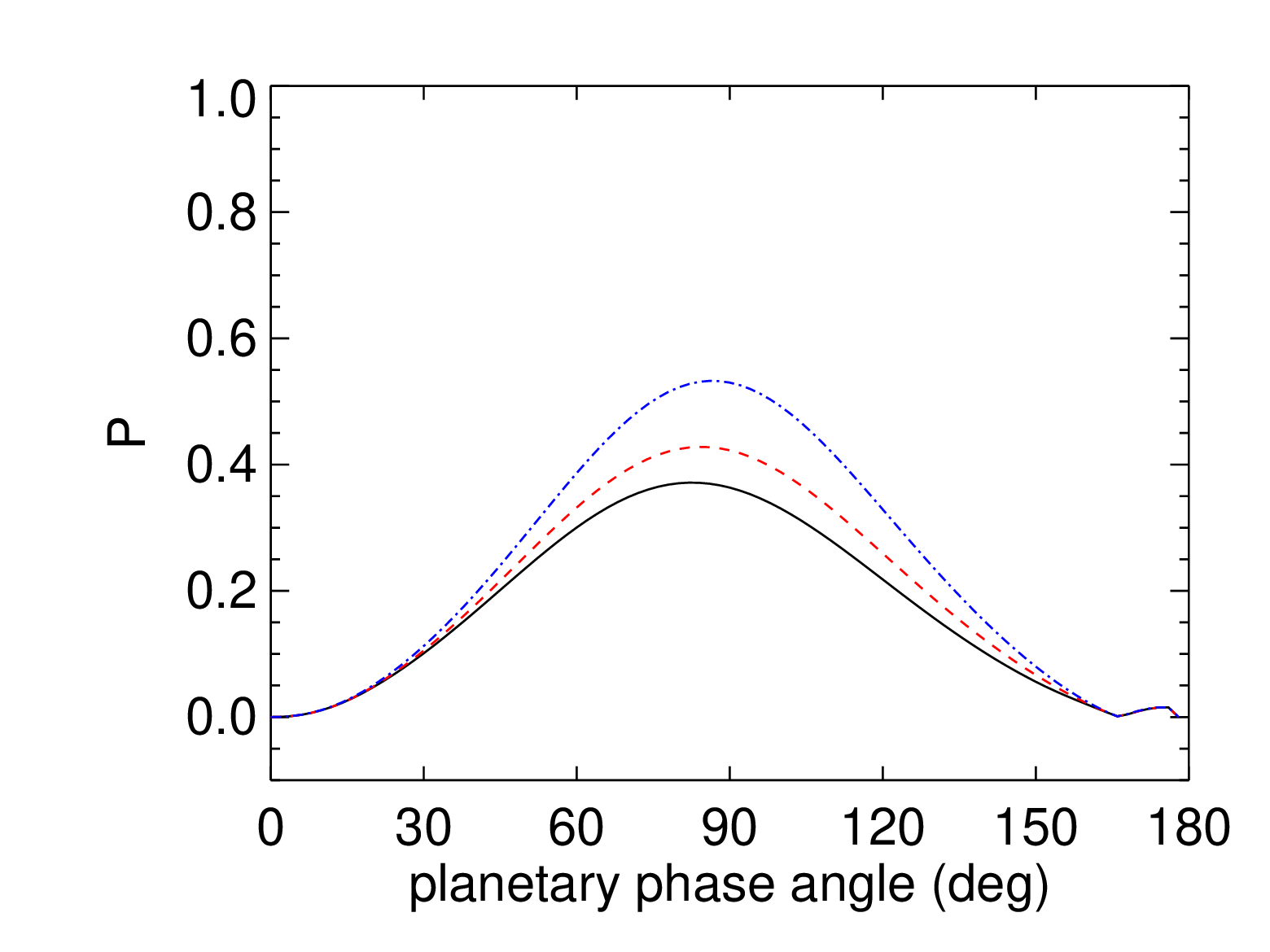}%{karalidi11b.ps}
\caption{$\pi F_\mathrm{n}$ and $P$ as functions of $\alpha$ for
  black, cloud--free planets with white, hemispherical caps that
  occupy 23$\%$ of the planet. The locations of the caps are:
  $0^\circ$ (black, solid line), indicating the caps are on the
  'north' and 'south' poles of the planet, $45^\circ$ (red, dashed
  line), and $90^\circ$ (blue, dashed--dotted line), with the caps on
  the 'eastern' and 'western' sides of the planetary disk.}
\label{fig:polcnt_inclin_fluxpol}
\end{figure}

Figure~\ref{fig:polcnt_inclin_fluxpol} shows $\pi F_\mathrm{n}$ and
$P$ for planets with caps covering 23\% of the disk for three
different locations of the caps. The three different flux phase
functions show the relatively small effects of different fractions of
the caps being visible at a given phase angle.  The polarisation phase
functions are all fairly symmetric around $\alpha=90^\circ$,
especially for the planet with its caps at the poles of the planet.

%-------------------------------------------------------------------------
% Figure 9: made with MX_PL_PARAL/N_RES_PAP2/DATA/POLAR/compare.pro
%-------------------------------------------------------------------------
\begin{figure}
\centering
\includegraphics[width=85mm]{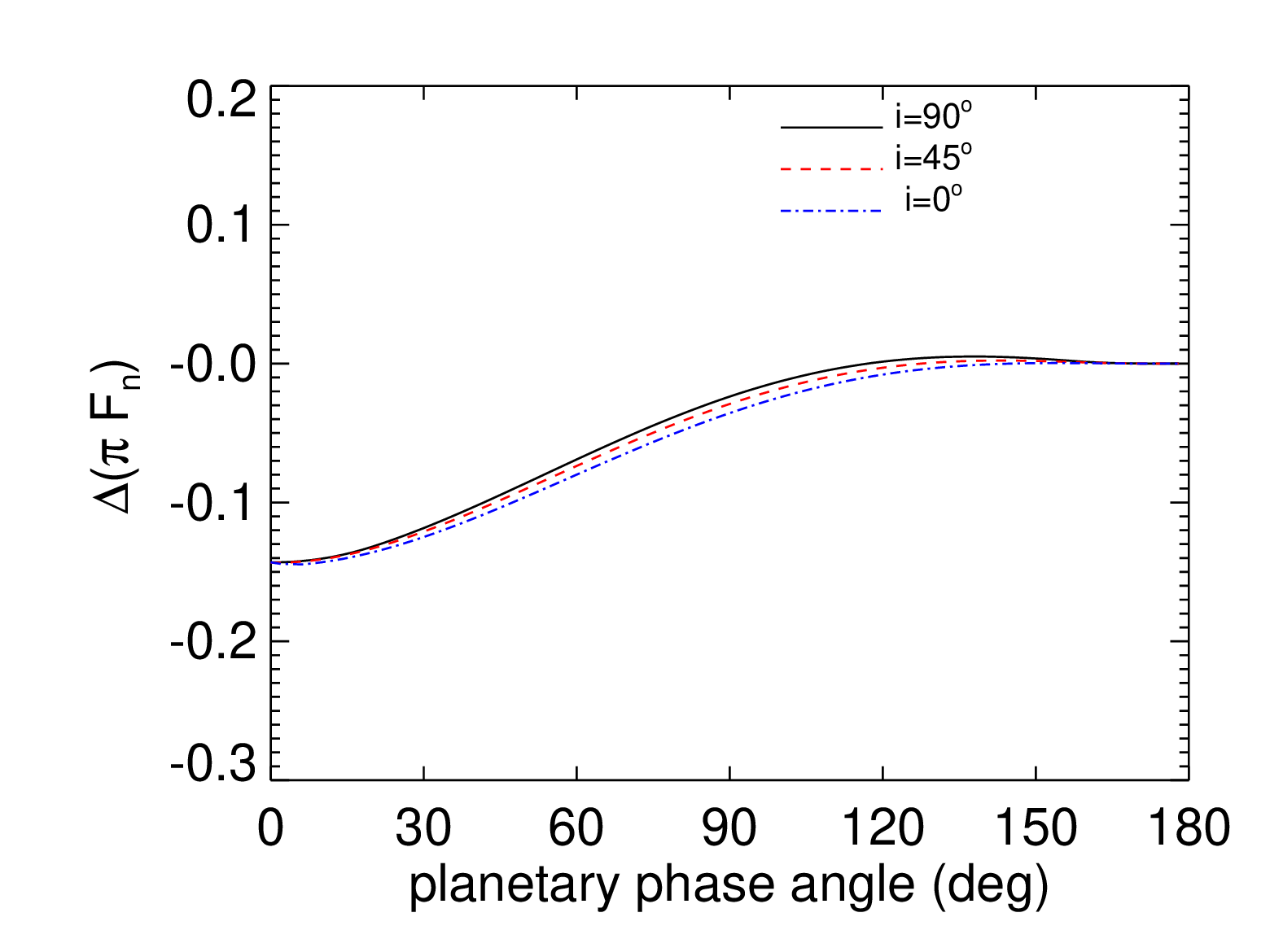}%{karalidi12a.ps}
\hspace{0.8cm}
\centering
\includegraphics[width=85mm]{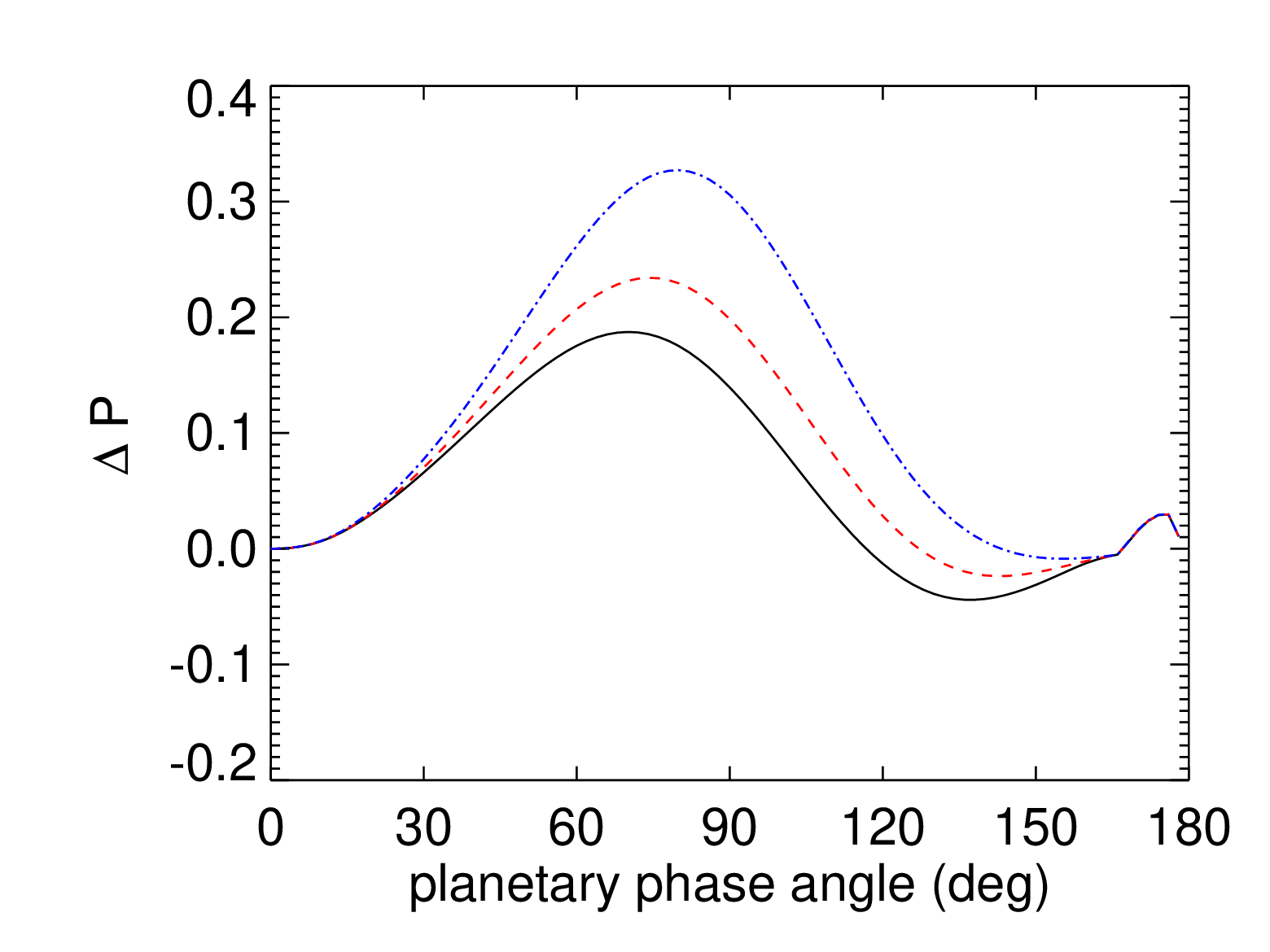}%{karalidi12b.ps}
\caption{$\Delta \pi F_\mathrm{n}$ and $\Delta P_\mathrm{s}$ between
  the phase functions of the capped planets of
  Fig.~\ref{fig:polcnt_inclin_fluxpol} as calculated using the HI-code
  and the HH-code.  The cap locations are: $0^\circ$ (black, solid
  line), $45^\circ$ (red, dashed line), and $90^\circ$ (blue,
  dashed--dotted line).}
\label{fig:polcnt_inclin_fluxpol_var}
\end{figure}

Figure~\ref{fig:polcnt_inclin_fluxpol_var} shows $\Delta \pi
F_\mathrm{n}$ and $\Delta P$ for the three model planets of
Fig.~\ref{fig:polcnt_inclin_fluxpol} as calculated using the HH-code
and the HI-code. As can be seen, the $\Delta \pi F_\mathrm{n}$ are
fairly independent of the caps' position angle. This implies that it
would be difficult to retrieve information about the position of the
caps from the flux alone.  Polarisation appears to be more sensitive
to the location of the caps.  In particular, the $\Delta P$ are
several percentage points depending on the location and on $\alpha$.
Around $\alpha=90^\circ$, where exoplanets have a relatively high
chance of being observed with direct detection methods, the
differences in $\Delta P$ are about 0.12 for the planet with the
eastern and western caps, implying that the accuracy of the
polarimetry should be larger than 0.10 to be able to establish their
existence.

%%%%%%%%%%%%%%%%%%%%%%%%%%%%%%%%%%%%%%%%%%%%%%%%%%%%%%%%%%%%%%%%%%%%%%%%%%
\section{Application to Earth--like planets}
\label{sec:sect_4}

In this section, we present flux and polarisation signals of planets
with horizontally inhomogeneous Earth--like surface coverages and
patchy cloud layers. In reality, the Earth exhibits a large variation
in atmospheric temperature and pressure profiles, cloud properties
(both on macro- and micro-scales) and surface properties.  To avoid
introducing too many variables, we assume a single temperature and
pressure profile across our model planet, the so--called mid--latitude
summer profile from \citep{mcclatchey72}, a single type of cloud
particles and cloud properties (optical thickness and vertical
distribution), and a surface that is covered by either ocean or sand
(a continent).

%-------------------------------------------------------------------------
\subsection{Surface inhomogeneities}
\label{sec.surf}

First, we present the signals of a planet with a sandy continent
surrounded by ocean. The continent extends between longitudes of $\pm
22^\circ$ and latitudes of $\pm 50^\circ$ (measured with respect to
the subobserver point).  The sand surface reflects Lambertian
(unpolarized and isotropically) with an albedo of 0.25. This albedo is
taken from the ASTER spectral library and should be representative for
a sand surface on Earth at $\lambda=0.55~\mu$m.  The ocean is
black. The planetary atmosphere is cloud--free, with a total gas
optical thickness of 0.1 (no absorption), which corresponds to a
wavelength of 0.55~$\mu$m.

\begin{figure}
\centering
\includegraphics[width=85mm]{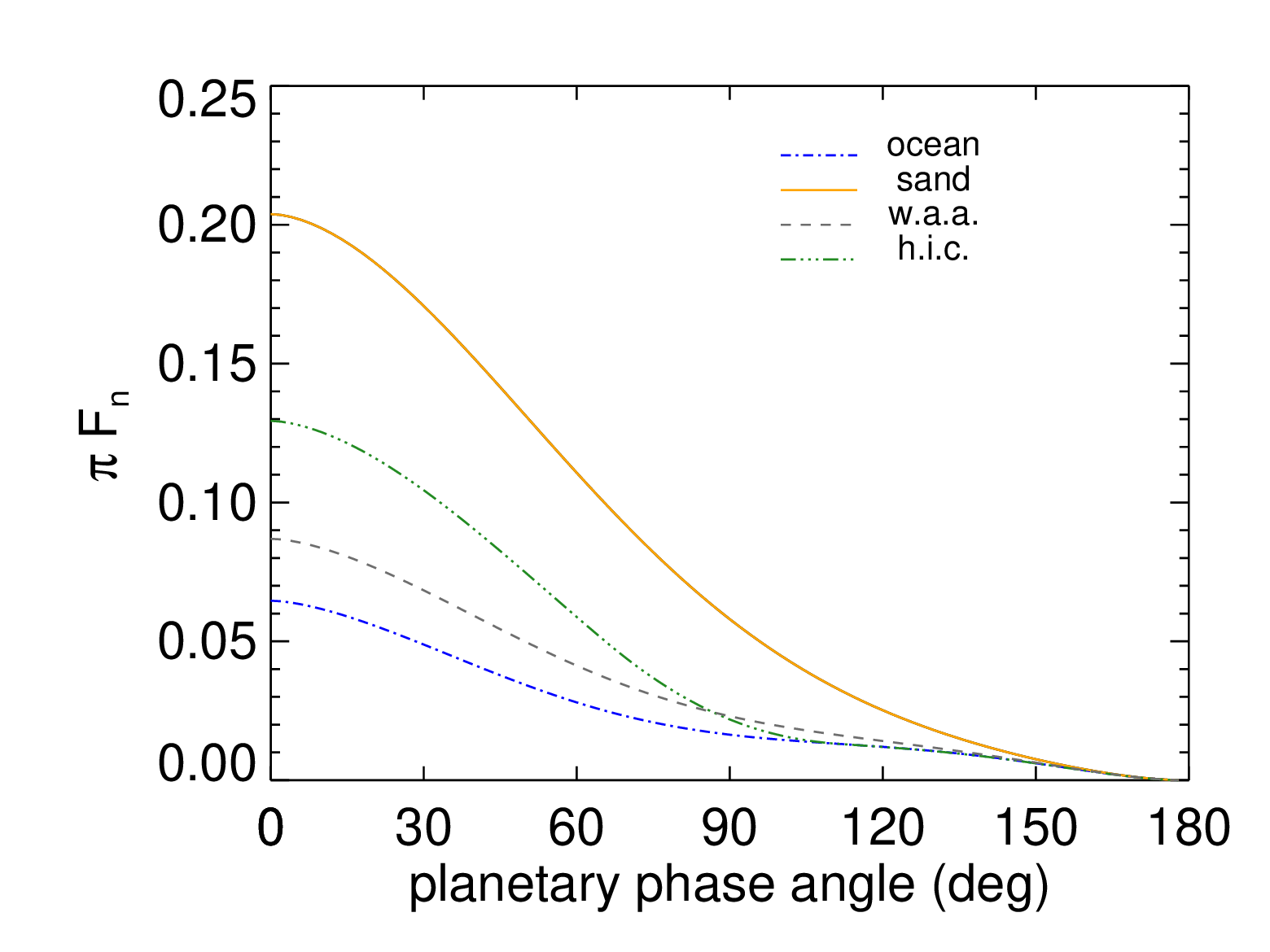}%{cent_cont_plot_f_16pc.ps}
\hspace{0.8cm}
\centering
\includegraphics[width=85mm]{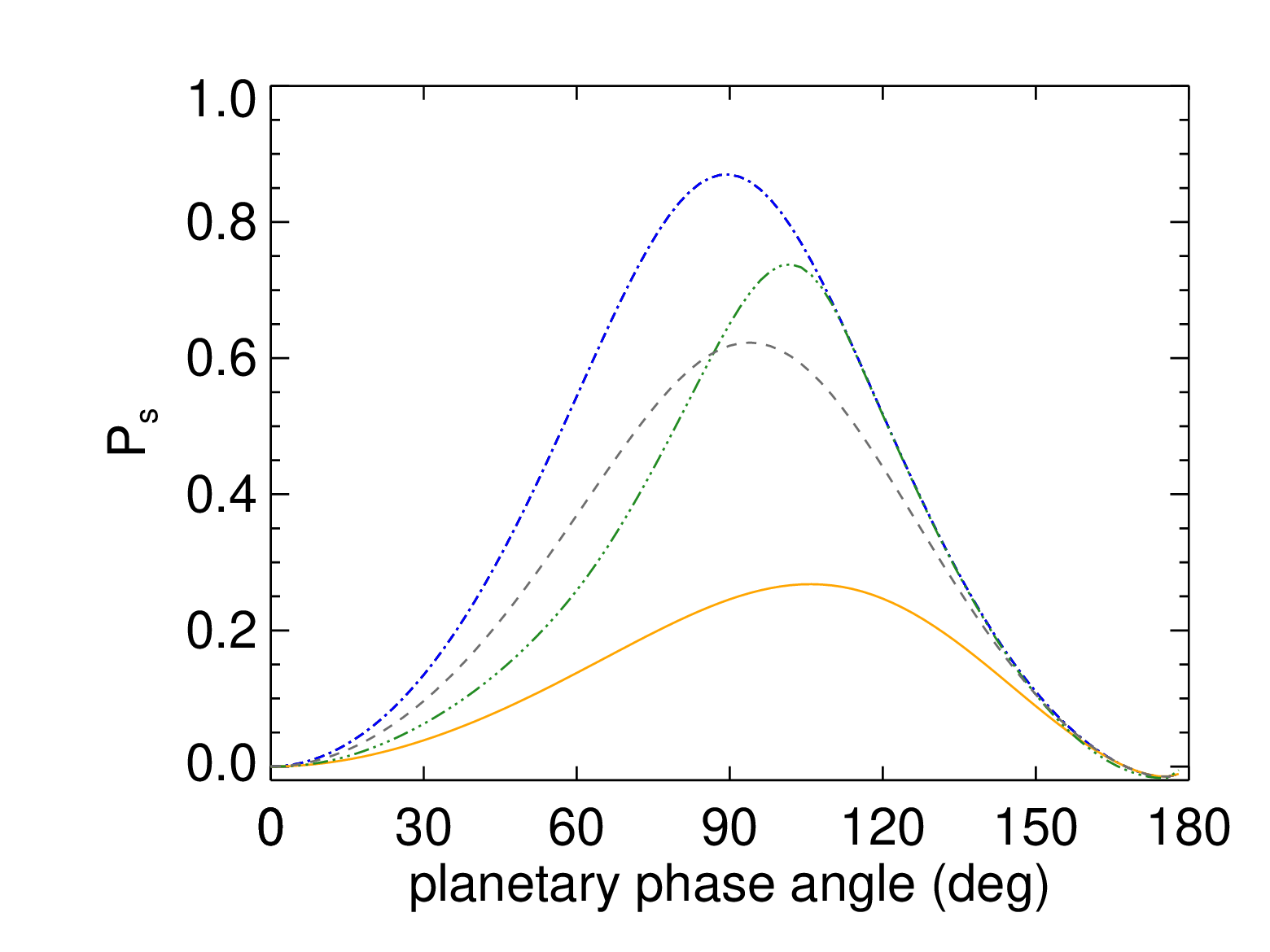}%{cent_cont_plot_p_16pc.ps}
\caption{$\pi F_\mathrm{n}$ and $P_\mathrm{s}$ as functions of
  $\alpha$ for a model planet covered by ocean and with a center
  continent of sand between $\pm22^\circ$ longitude and $\pm50^\circ$
  latitude.  (green, dashed--triple-dotted line). Also plotted: the
  functions for a homogeneous ocean planet (blue, dashed--dotted
  line), a homogeneous sand planet (orange, solid line) and a weighted
  sum of these with $\sim$16\% sand and $\sim$84\% ocean (gray, dashed
  line).}
\label{fig:cent_cont}
\end{figure}

Figure~\ref{fig:cent_cont} shows $\pi F_\mathrm{n}$ and $P_\mathrm{s}$
as functions of $\alpha$ for the model planet as calculated with our
HI-code and the HH-code (combined with the weighted averages method),
and, to compare, for model planets that are completely covered by sand
or ocean.  Both in $\pi F_\mathrm{n}$ and in $P_\mathrm{s}$, there are
significant differences between the signals from the HI-code and the
HH-code, depending on the phase angle.  At $\alpha=0^\circ$, the
difference $\Delta \pi F_{\rm s}$ is $\sim$~30\%. With increasing
phase angle, the fraction of the continent on the planet's nightside
increases, and $\pi F_\mathrm{n}$ calculated with the HI-code
decreases faster than that calculated with the HH-code.  The
polarisation phase function calculated using the HI-code is more
asymmetric than that calculated using the HH-code. The largest
absolute difference between the two curves is $\sim$0.14 (14\%) at
$\alpha=106^\circ$.
%-------------------------------------------------------------------------
% Figure 10:
%-------------------------------------------------------------------------

Incidentally, for this planet, the flux and polarisation signals
calculated using the HI-code and the HH-code are virtually equal at
$\alpha=90^\circ$. At this phase angle, flux and polarisation
observations would thus not help to establish the existence of the
continent. However, interpreting flux observations at smaller (larger)
phase angles using the HH-code would result in an overestimation
(underestimation) of the coverage with sand, while interpreting only
polarisation observations at smaller (larger) phase angles using the
HH-code would result in an underestimation (overestimation) of the
coverage with sand. The interpretation of the combined flux and
polarisation observations using the HH-code would fail and thus reveal
that assuming a homogeneous mixture of sand and ocean pixels is not
realistic.

%-------------------------------------------------------------------------
\subsection{Atmospheric inhomogeneities}
\label{sec.clouds}

The next model planet has a black surface, a gaseous atmosphere with
optical thickness of 0.1 (no absorption), and a patchy cloud layer
with an optical thickness of 2.0. The clouds are composed of liquid
water particles described in size by the standard distribution of
\citet{hansentravis74}, with an effective radius of 2.0~$\mu$m and an
effective variance equal to 0.1 (model A particles of
\citet[][]{karalidi11}). The single scattering albedo of our particles
at 0.55 $\mu$m is 0.999534. For a detailed description of the single
scattering properties of our cloud particles see
\citet[][]{karalidi11}. The cloud layer is patchy (the clouds are
composed of fully cloud covered pixels that cluster in random manner
across the planetary surface) in the horizontal direction, but its
vertical extension is the same all over the planet, namely from 3 km
to 4 km in Earth's atmosphere (279 K to 273 K).

%-------------------------------------------------------------------------
% Figure 11: /data/theodora/MIXED_PLANETS_MODIFIED/compare_cldeck.pro
%-------------------------------------------------------------------------
\begin{figure}
\centering
\includegraphics[width=85mm]{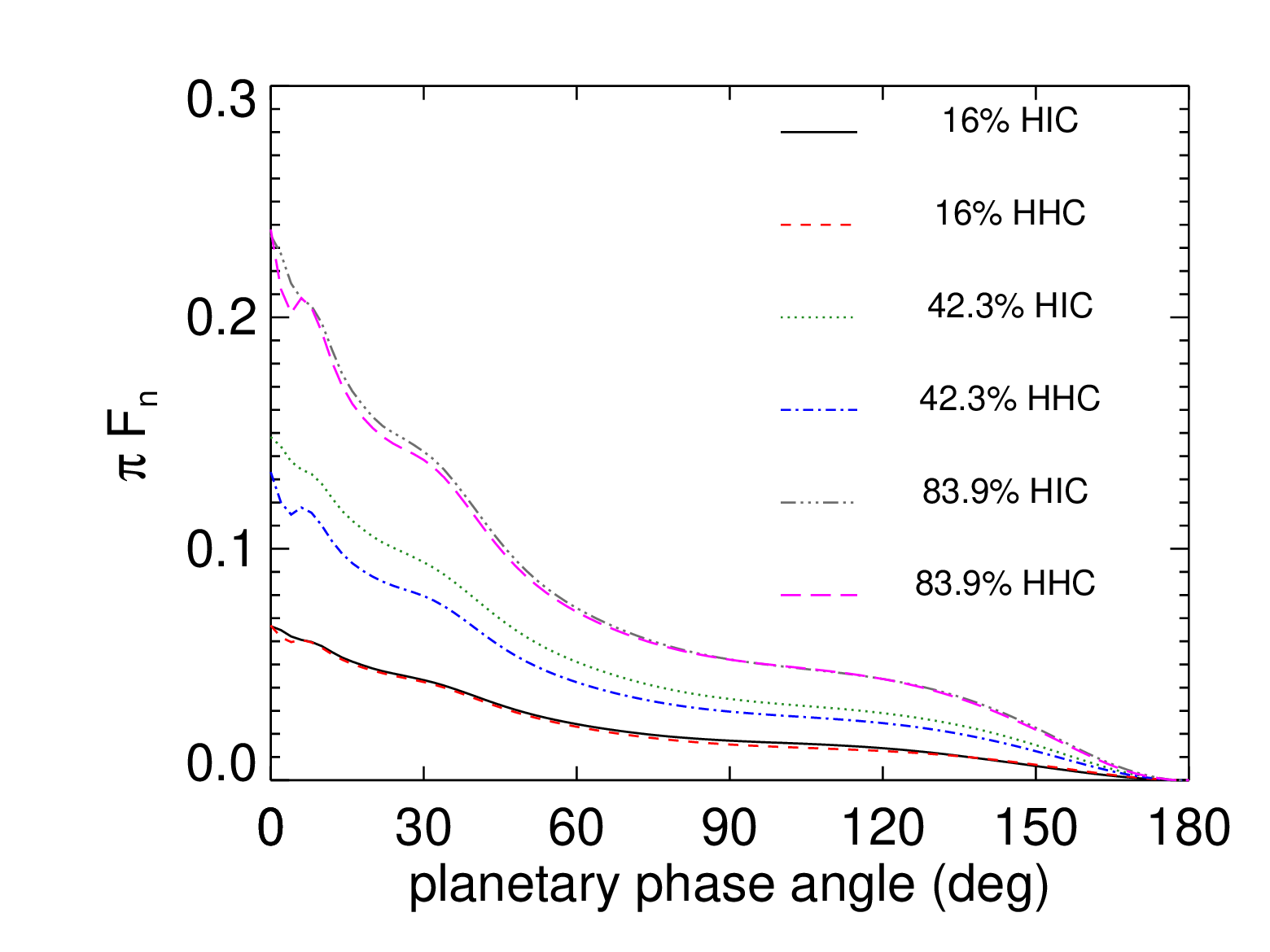}%{various_cloud_coverages_f.ps}
\hspace{0.8cm} \centering
\includegraphics[width=85mm]{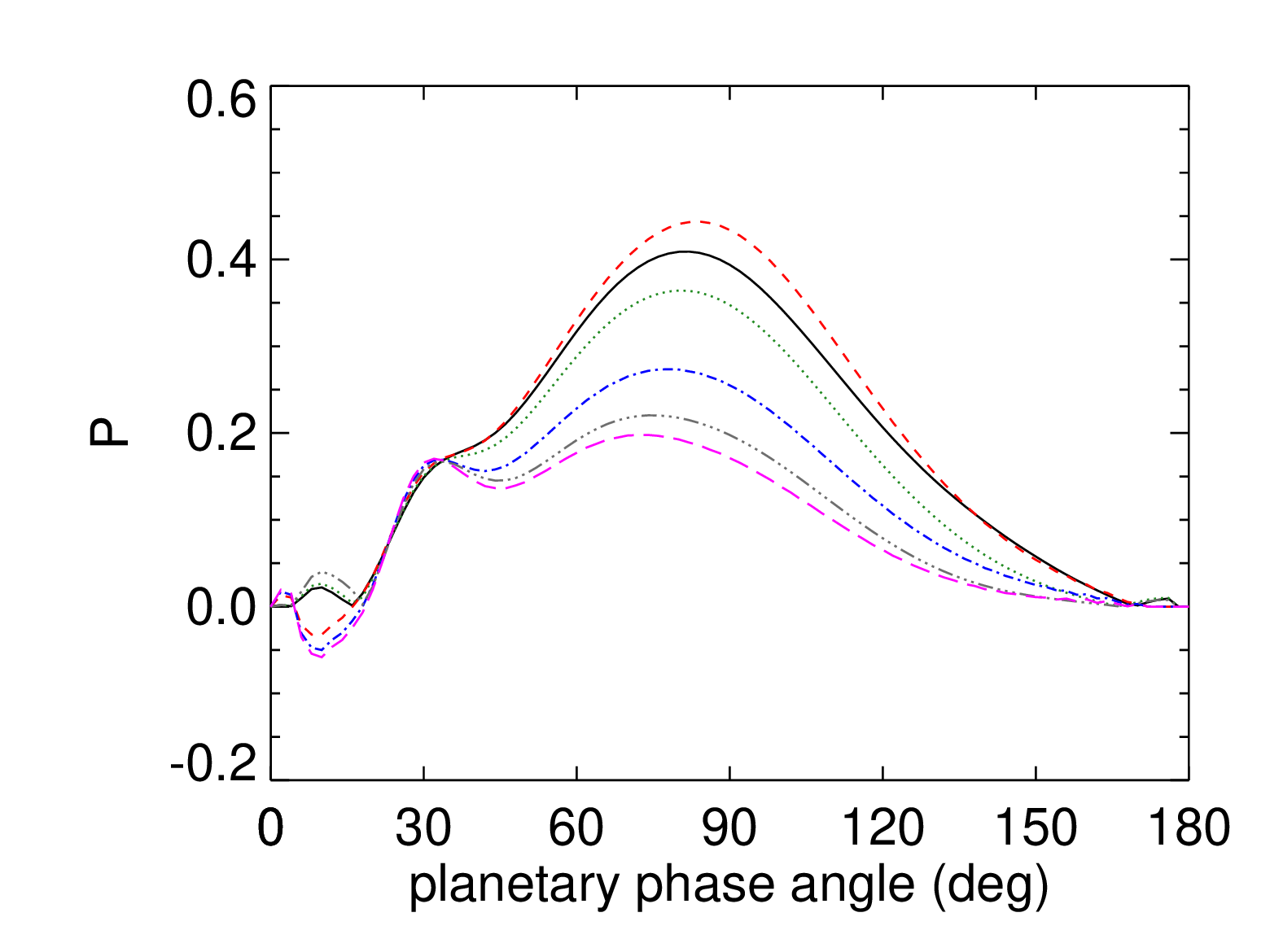}%{various_cloud_coverages_p.ps}
\caption{$\pi F_\mathrm{n}$ and $P_\mathrm{s}$ as functions of
  $\alpha$ for a model ocean planet with patchy clouds
  that cover 16\%, 42.3\%, or 83.9\% of the planet.
  The curves as calculated using the HH-code and the weighted
  sum approximation are also shown.}
\label{fig:var_cl_cvs}
\end{figure}

Figure~\ref{fig:var_cl_cvs} shows $\pi F_\mathrm{n}$ and $P$ as
functions of phase angle $\alpha$ of the model planet for different
cloud coverages as calculated using the HI- and the HH-code.  The flux
phase functions for the different cloud coverages have different
strengths, but very similar shapes. The fase functions as calculated
using the HH-code have slightly sharper features near
$\alpha=0^\circ$.  Note that the bump in the curves near
$\alpha=30^\circ$ is the signature of the primary rainbow \citep[see
  e.g.][]{karalidi11}: light that is scattered once by the cloud
particles.  As expected, the difference between the fluxes calculated
with the two codes are small for small and large percentages of cloud
coverage.  For an intermediate, almost Earth--like, coverage of
42.3\%, $\Delta \pi F_\mathrm{n}$ is as large as 15\% at
$\alpha=0^\circ$.  In this case, using the HH-code to interpret the
flux reflected by the patchy cloudy planet near $\alpha=0^\circ$ would
yield a cloud coverage of $\sim$53\%.

The polarisation phase functions clearly show the different
contributions of the light that is Rayleigh scattered by the gas
molecules above and in particular between the patches of clouds (the
strong values around $\alpha=90^\circ$) and that of the light that is
scattered by the cloud articles (the primary rainbow).  Clearly, the
larger the coverage of the clouds, the smaller the contribution of the
purely Rayleigh scattered light to the total signal.  Adding clouds to
our model planet decreases the amount of purely Rayleigh scattered
light, and thus decreases $P$ around $\alpha=90^\circ$.  The strength
of the primary rainbow in $P$ is insensitive to the cloud coverage
because adding clouds does not change the fraction of multiply
scattered light within the clouds (the clouds have the same optical
properties), which mostly determines the strength of the rainbow on
these planets. The Rayleigh scattering polarisation maximum around
$\alpha=90^\circ$ does influence the contrast of the rainbow feature:
for low cloud coverages it forms a shoulder on the Rayleigh scattering
maximum, while for high coverages, it is a local maximum.

For small and large percentages of cloud coverage, the differences in
the polarisation phase functions due to using either the HI- or the
HH-code are at most a few percent points around $\alpha=90^\circ$
($\sim$-4\% for 16\%, and $\sim$3\% for 83.9\% coverage).  The
differences are largest for the intermediate cloud coverage: about
10\% for 42.3\% coverage.  In this case, using the HH-code to
interpret the polarisation reflected by the patchy cloudy planet near
$\alpha=90^\circ$, would yield a cloud coverage of $\sim$25\%.  The
degree of polarisation calculated with the HI-code, and thus $\Delta
P$ too, depends not only on the cloud coverage, but also on the
locations of the clouds across the planet. Not surprisingly, this
sensitivity is highest for intermediate cloud coverages.

%-------------------------------------------------------------------------
\subsection{Atmospheric and surface inhomogeneities}\label{sec:mixedpl}

\begin{figure}
\centering
\includegraphics[width=85mm]{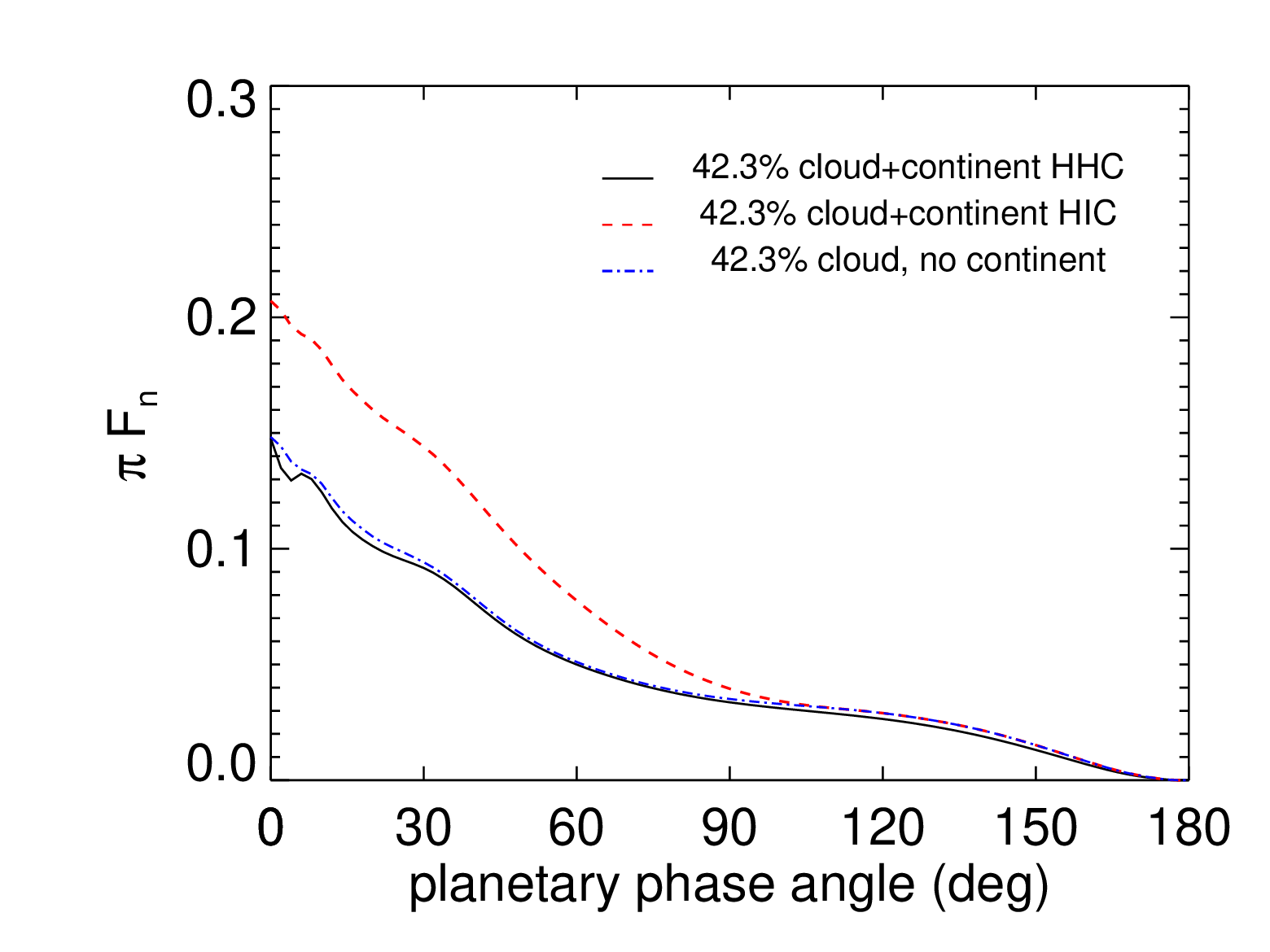}%{mixed_atm_srf_F_comp.ps}
\hspace{0.8cm} \centering
\includegraphics[width=85mm]{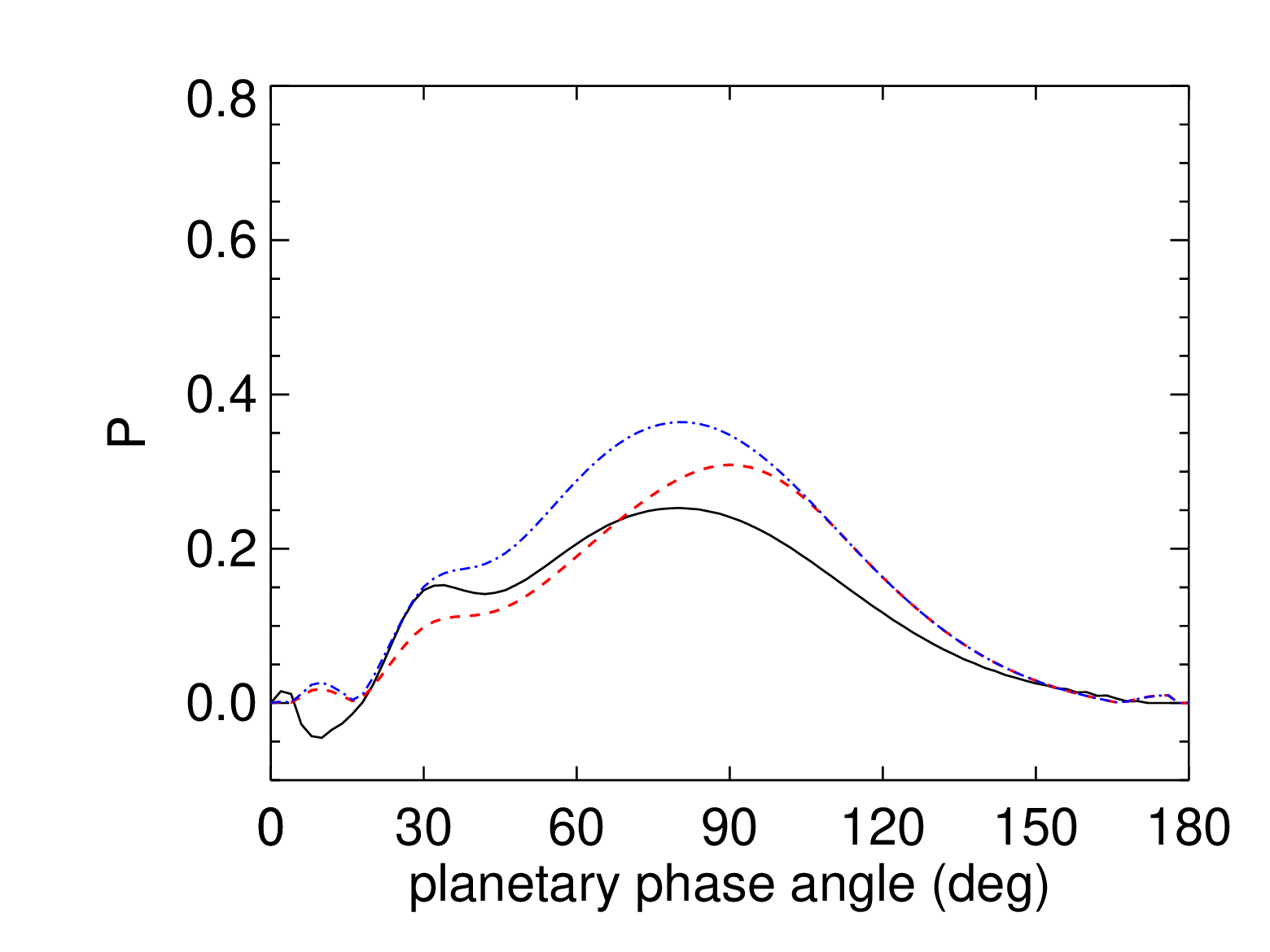}%{mixed_atm_srf_P_comp.ps}
\caption{$\pi F_\mathrm{n}$ and $P_\mathrm{s}$ as functions of
  $\alpha$ for a model ocean planet with a central sandy continent and
  42.3\% cloud coverage as calculated using the HI-code (red, dashed
  line) and the HH-code and the weighted sum approximation (black,
  solid line).  For comparison, the curves for a model ocean planet
  without the continent and with 42.3\% cloud coverage are also shown
  (blue, dashed--dotted line).}
\label{fig:combined_variations}
\end{figure}

Finally, we present the flux and polarisation signals of planets with
patchy clouds (see Sect.~\ref{sec.clouds}), and a sandy continent in
the middle of a black ocean (see Sect.~\ref{sec.surf}).
Figure~\ref{fig:combined_variations} shows $\pi F_\mathrm{n}$ and $P$
as functions of $\alpha$ for 42.3\% cloud coverage. The phase
functions as calculated using the HH--code and the HI--code results
for a cloudy planet without a continent are also shown.

The continent strongly increases $\pi F_\mathrm{n}$. The increase will
usually depend on the location of the clouds and the continent.  In
this case, the continent is located in the middle of the planetary
disk, and thus has a large influence. Since only a small fraction
($\sim$ 8\%) of the pixels on the disk contain sand, the flux
calculated with the HH-code is only marginally higher than that
calculated with the HH-code and without a continent (see
Fig.~\ref{fig:var_cl_cvs}).  Inversely, if the HH-code would be used
to interpret the flux signal of the cloudy planet with the continent,
a 55\% continental coverage and 24\% clouds would be found.

%-------------------------------------------------------------------------
% Figure 12:  MADE WITH /data/theodora/MIXED_PLANETS_INT/comp_loc_cont.pro
%-------------------------------------------------------------------------

The presence of the Lambertian reflecting continent below the clouds
decreases the polarisation phase function and shifts the maximum $P$
towards larger $\alpha$, when compared to the polarisation phase
function of the cloudy ocean planet. The primary rainbow is still
visible in the phase function, albeit more like a shoulder than a
local maximum.

Using polarimetry only, a straightforward fit to the polarisation
phase function with the HH-code would estimate the continental
coverage at about 17\% and the cloud coverage at 24\%.  When both $\pi
F_\mathrm{n}$ and $P$ are taken into account the best fit is acquired
for the case of 22\% continental coverage and 19\% cloud coverage.

%            ocean            cloud        cont             cl+cnt
%flux only: 0.14000000      0.18000000      0.55000000     0.060000000 
%pol   0.82000000     0.060000000      0.17000000      0.18000000 
%both continent  0.18000000     0.020000000      0.79000000     0.010000000 
 %             0.060000000      0.58000000      0.11000000      0.19000000
%%%%%%%%%%%%%%%%%%%%%%%%%%%%%%%%%%%%%%%%%%%%%%%%%%%%%%%%%%%%%%%%%%%%%%%%%%
%\clearpage

\section{Summary and conclusions}
\label{sec:sect_5}

We have presented a numerical code that can be used to calculate disk
integrated total and polarised fluxes of starlight that is reflected
by horizontally inhomogeneous exoplanets, e.g. planets that are
covered by oceans and continents, and/or overlaid by a patchy cloud
deck.

For most types of model planets, our code for horizontally
inhomogeneous planets (the HI-code) is computationally much more
expensive than the code for horizontally homogeneous planets by
\citet{stam06} combined with the so--called weighted sum approximation
to simulate fluxes of horizontally inhomogeneous planets (the
HH-code). In the latter method, total and polarized flux signals of
different horizontally homogeneous planets are summed using weighting
factors depending on which fraction of the illuminated and visible
part of the horizontally inhomogeneous planet is represented by each
type of planet.  Only for planets with a large variation of horizontal
inhomogeneities, such as a large number of different surface and cloud
coverages, the computing time for the HH-code approaches that for the
HI-code.

The main advantage of using the HI-code instead of the HH-code is
obviously the ability to simulate signals of horizontally
inhomogeneous planets. Other advantages of the HI-code are that since
every pixel on the planet is treated separately, it allows taking into
account effects of e.g. the non-sphericity of a planet, shadowing by
planetary rings, and a spatial extension of the illuminating source.

Since for most model planets, our HI-code consumes much more computing
time than the HH-code \citep[][]{stam06}, it is interesting to
investigate the influences of horizontal inhomogeneities on the total
and polarized fluxes of starlight that is reflected by a planet.  For
three types of horizontally inhomogeneous planets covered by black and
white surface pixels and overlaid by a gaseous atmosphere, we have
calculated the fluxes and the degree of polarisation as functions of
the planetary phase angle $\alpha$ and compared them with results from
the HH-code assuming the same percentage of black and white surface
pixels and the same model atmosphere.  Horizontal inhomogeneities can
leave significant traces in both the reflected total flux $\pi
F_\mathrm{n}$ and the degree of linear polarisation. However, while
horizontal inhomogeneities appear to mostly influence the total amount
of reflected flux and not so much the shape of the planet's flux phase
functions, they can strongly change the planet's polarisation phase
functions in shape and strength. Indeed, fitting a planet's
polarisation phase function using the HH-code would yield very
different fractions of disk coverage (up to several tens of percent).

We also used the HI-code to calculate flux and polarisation signals of
Earth--like planets with surface and/or patchy clouds.  These
calculations confirmed that the shape of a planet's flux phase
function is fairly independent of the horizontal
inhomogeneities. Obviously, the absolute values of the flux phase
function does depend on the inhomogeneities, but it will also depend
on the radius of the planet (in this paper we assumed a planetary
radius equal to one), which thus would have to been known accurately
to fit the flux phase function. The polarisation phase function
appears to be rather sensitive to the inhomogeneities, both its
absolute values and the shape of the curve (e.g. location maximum
value).  Because the degree of polarisation is a ratio, it is
independent of the radius of a planet. The flux and polarisation
signals of planets are wavelength dependent \citep[see
  e.g.][]{stam08}, therefore the observability of horizontal
inhomogeneities will also depend on the wavelength.  In particular, at
short wavelengths, the gas optical thickness is larger than at longer
wavelengths, and will thus hamper the observations of the
surface. Future studies could focus on which spectral bands should be
combined to optimize retrieval schemes.

In the presence of liquid water clouds, the strength of the primary
rainbow in the flux phase functions of Earth-like planets increases
with increasing cloud coverage. Whether or not it would be detectable
depends strongly on the sensitivity and stability of the observing
instrument.  In the polarisation phase function, the strength of the
rainbow is almost independent of the cloud coverage as long as the
planet's surface is very dark (i.e. does not add too much unpolarized
flux to the total signal). A bright surface, such as a sandy
continent, decreases the strength of the rainbow feature.

Properly accounting for horizontal inhomogeneities appears to
significantly influence reflected fluxes and polarisation signals, and
should eventually be applied to interpret observations of horizontally
inhomogeneous exoplanets or e.g. observations of Earth-shine
\citep[][]{sterzik12}. The HH-code (horizontally homogeneous planets
combined with the weighted sum approximation), however, is still a
strong tool for simulating signals to be used for the design and
optimization of exoplanet observations, because its simulations cover
the range of total and polarized fluxes that we can expect to observe.
Because flux and polarisation phase functions have different
sensitivities to the inhomogeneities, a combination of flux and
polarisation observations would help to retrieve the actual planetary
parameters.

Our results, and especially the differences between the fluxes and
degrees of polarization of the reflected starlight, indicate which
accuracies should be reached with flux and/or polarization
observations in order to detect horizontal inhomogeneities on a
planet. As such they can drive the design of instruments for exoplanet
characterization.  Assuming a super--Earth exoplanet (with a radius of
1.5~r$_\oplus$) orbiting a Sun--like star at 1 pc from the observer at
a 40-meter telescope (such as the E-ELT), a back-of-the-envelope
calculation shows that a few nights of integration time (a total of 20
hours) could yield an accuracy of 10$^{-3}$. The phase angle of a
planet in an Earth--like orbit would change by $\sim$3$^\circ$ during
that time.  For this calculation we ignored the influence of stellar
background light on the observation, and we didn't include any actual
instrument parameters, such as spectral bandwidths. Until results such
as ours are combined with a realistic instrument and telescope
simulator, the integration time estimate should thus be considered as
a rough value.

%%%%%%%%%%%%%%%%%%%%%%%%%%%%%%%%%%%%%%%%%%%%%%%%%%%%%%%%%%%%%%%%%%%%%%%%%%
% Bibliography
%%%%%%%%%%%%%%%%%%%%%%%%%%%%%%%%%%%%%%%%%%%%%%%%%%%%%%%%%%%%%%%%%%%%%%%%%%
%\bibliographystyle{aa}
%{%\small
%\bibliography{references}
%}

%%%%%%%%%%%%%%%%%%%%%%%%%%%%%%%%%%%%%%%%%%%%%%%%%%%%%%%%%%%%%%%%%%%%%%%%%%
% APPENDICES
%%%%%%%%%%%%%%%%%%%%%%%%%%%%%%%%%%%%%%%%%%%%%%%%%%%%%%%%%%%%%%%%%%%%%%%%%%
\clearpage
\begin{appendices}

%%%%%%%%%%%%%%%%%%%%%%%%%%%%%%%%%%%%%%%%%%%%%%%%%%%%%%%%%%%%%%%%%%%%%%%%%%
\section{Testing our numerical code}
\label{appendix_A}

Here, we present the results of testing our numerical code for 
horizontally inhomogeneous planets (from hereon the HI-code).
We have tested our code by comparing it to results of the
code for horizontally homogeneous planets \citet{stam06}
(from hereon the HH-code).

%-------------------------------------------------------------------------
\subsection{Horizontally homogeneous planets}
\label{sect_test_homogeneous}

For the first comparison between our HI-code and the HH-code, we used
both codes to calculate the light reflected by a model planet with a
black surface and a gaseous, Rayleigh scattering atmosphere with a
total optical thickness of about~0.1. For the HI-code, the planet was
divided into pixels of $2^\circ \times 2^\circ$ (latitude $\times$
longitude).

\begin{figure}
\centering
\includegraphics[width=85mm]{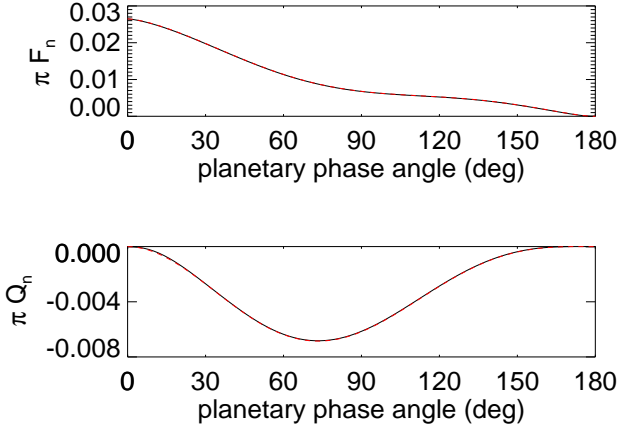}%{karalidi1.ps}
\caption{Normalised reflected fluxes $\pi F_\mathrm{n}$ and $\pi
  Q_\mathrm{n}$ as functions of the planetary phase angle $\alpha$ for
  a black planet with a gaseous atmosphere as calculated using
  the HH-code (solid lines) and the HI-code (dashed lines).
  The solid and dashed lines coincide.}
\label{fig:first_comp_blsurf}
\end{figure}

\begin{figure}
\centering
\includegraphics[width=85mm]{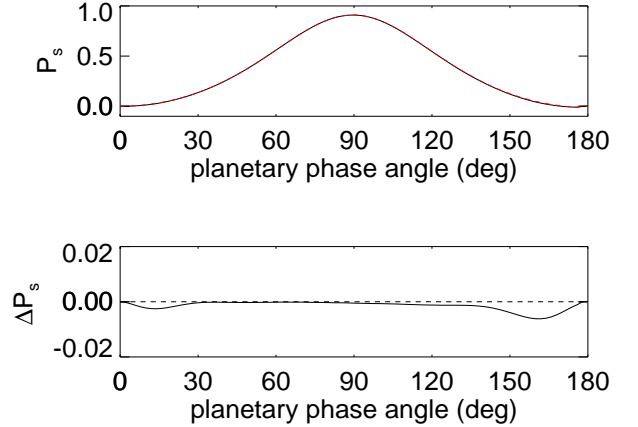}%{karalidi2.ps}
\caption{The degree of polarisation $P_\mathrm{s}$ corresponding to
  the fluxes shown in Fig.~\ref{fig:first_comp_blsurf} (the two lines
  coincide) and the absolute difference between the two lines.}
\label{fig:first_comp_p_blsurf}
\end{figure}

Figure~\ref{fig:first_comp_blsurf} shows the excellent agreement
between the reflected fluxes $\pi F_\mathrm{n}$ and $\pi Q_\mathrm{n}$
as functions of phase angle $\alpha$ calculated using the two codes
(because the planet is mirror--symmetric with respect to the reference
plane, $\pi U_\mathrm{n}= 0$). The absolute difference between the
fluxes calculated by the two codes is smaller than
$10^{-5}$. Figure~\ref{fig:first_comp_p_blsurf} shows the degree of
polarisation $P_\mathrm{s}$ ($= -Q/F$) (see Eq.~\ref{eq:signedP})
calculated using both codes, and the absolute difference $\Delta
P_\mathrm{s}$ between the curves.  $\Delta P_\mathrm{s}$ is largest
around small (15$^\circ$) and large (160$^\circ$) phase angles, which
is due to the size of the pixels.  Since in the HI-code, we use one
set of angles $\theta_0$, $\theta$, $\Delta \phi$, and $\beta_2$ for
each pixel, the larger a pixel, the less this set of angles represents
the range of angles across the pixel. The error is largest for pixels
along the planetary limb and terminator, and thus for the large and
small phase angles (at the small angles, the degree of polarisation in
the centre of the disk is close to zero, making the contribution of
the limb pixels thus significant).  At $\alpha=0^\circ$ and
180$^\circ$, $P_\mathrm{s}$ of the planet is zero and thus $\Delta
P_\mathrm{s}$, too.
%-------------------------------------------------------------------------
% Appendix: Figure 1:
%-------------------------------------------------------------------------

%-------------------------------------------------------------------------
% Appendix: Figure 2:
%-------------------------------------------------------------------------

For a model planet with a gaseous atmosphere, the differences between
the results of the two codes appear to be fairly independent of the
albedo of a (Lambertian reflecting) surface: with a white surface, the
absolute differences in $\pi F_\mathrm{n}$ and $\pi Q_\mathrm{n}$, and
$\Delta P_\mathrm{s}$ are similar to those with a black surface.  For
a model planet with a horizontally homogeneous cloud layer of optical
thickness 1, composed of the model A cloud particles of
\citet{karalidi11}, and a surface albedo of 0.1, the absolute
difference in the fluxes increased slightly to a maximum value of
5$\times$10$^{-5}$ and the maximum $\Delta P_\mathrm{s}$ increased to
$\sim 0.0062$. These differences didn't change significantly when
other (e.g. larger) cloud particles were used.

The comparison with the HH-code shows that the integration across the
disk by our HI-code is accurate enough for application to horizontally
homogeneous planets, provided small enough pixels are used across the
disk.

%-------------------------------------------------------------------------
\subsection{Horizontally inhomogeneous planets}
\label{sect_test_inhomogeneous}

For a second test of our HI-code, we used model planets with surfaces
covered by two types of $2^\circ \times 2^\circ$ pixels that alternate
in longitude and latitude (they thus look like chess boards). The
equator of each planet coincides with the planetary scattering plane.
The planetary atmospheres are gaseous, Rayleigh scattering and have an
optical thickness of 0.1.  While our HI-code can handle these types of
planets, the HH-code cannot. Therefore, as in \citet{stam08}, weighted
sums of flux vectors reflected by horizontally homogeneous planets are
used to approximate the flux vectors of horizontally inhomogeneous
planets. The flux vector of a planet covered by $J$ different types of
pixels (with a different atmosphere and/or surface) is thus calculated
using
%------------------------------------
\begin{equation}
   \pi \vec{F}(\alpha)= \sum_{j=1}^{J} w_j \hspace*{0.1cm} 
       \pi \vec{F}_j(\alpha) \hspace*{0.5cm} {\rm{with}} 
       \hspace*{0.2cm} \sum_{j=1}^{J} w_j = 1
\label{eq:weightav}
\end{equation}
%------------------------------------
with $\pi F_j$ the flux vector of starlight reflected by a planet that
is completely covered by type $j$ pixels, and with $w_j$ the fraction
of type $j$ pixels on the horizontally inhomogeneous planet.

\begin{figure}
\centering
\includegraphics[width=85mm]{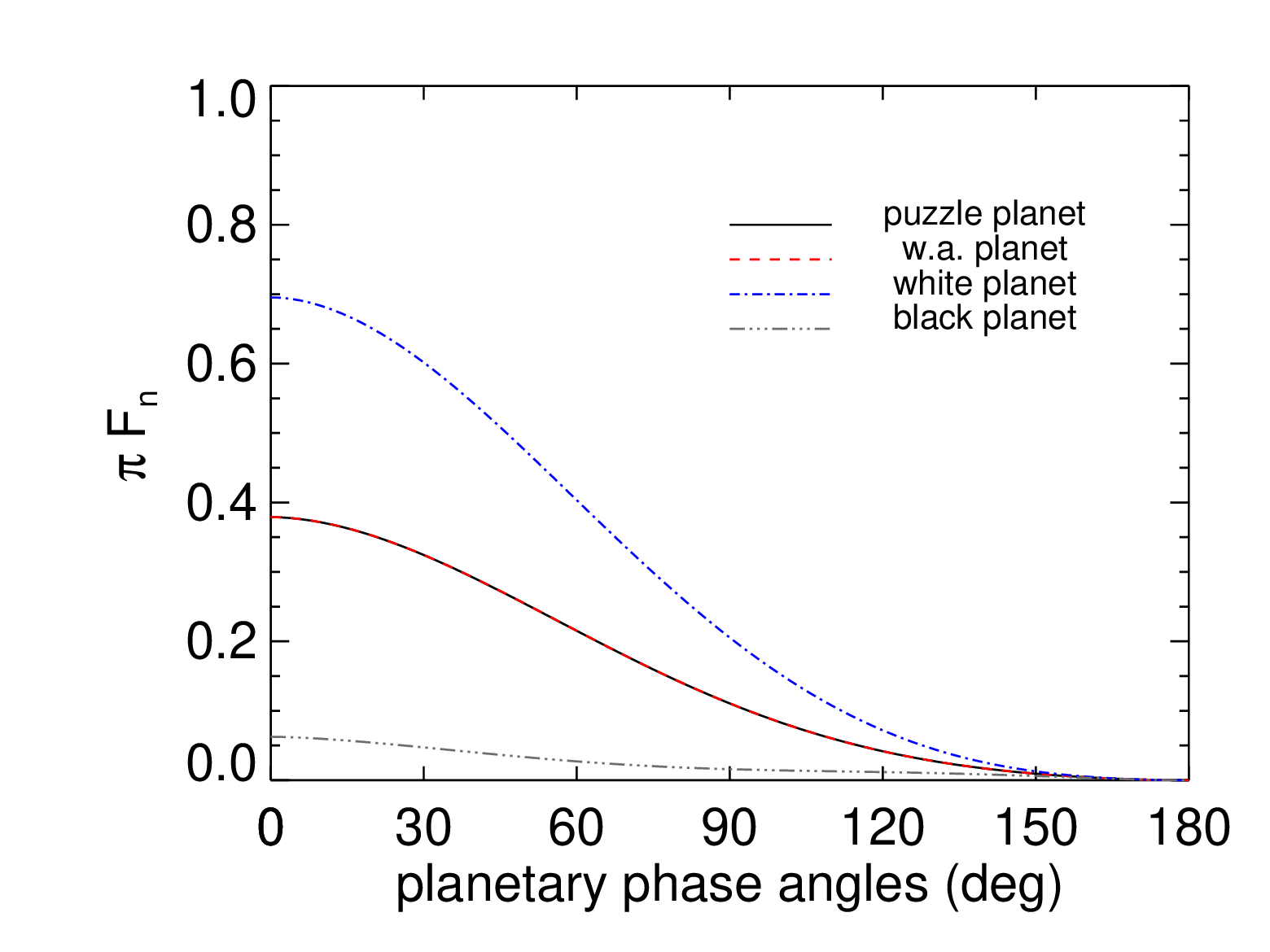}%{karalidi3a.ps}
\hspace{0.8cm}
\centering
\includegraphics[width=85mm]{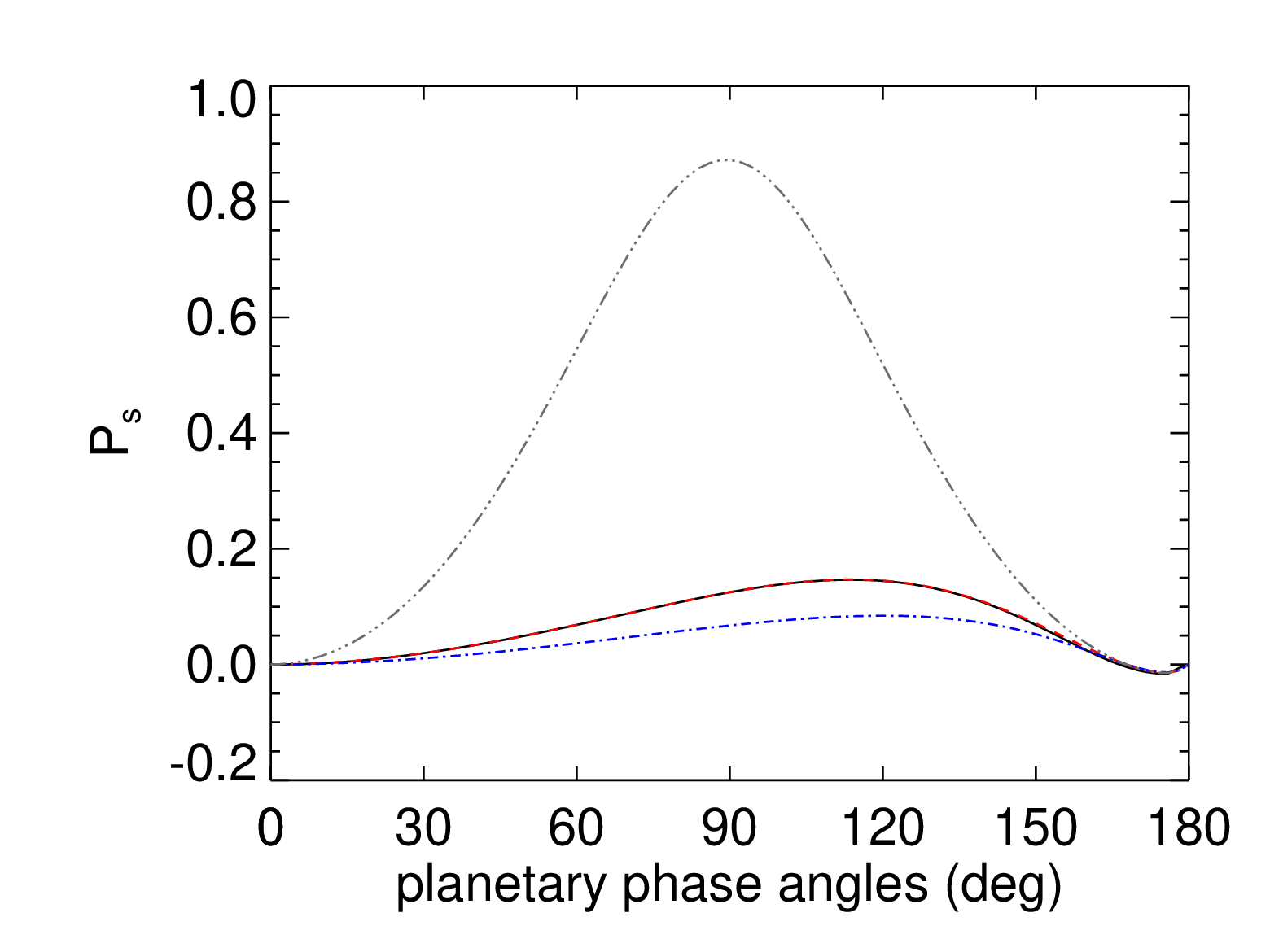}%{karalidi3b.ps}
\caption{$\pi F_\mathrm{n}$ and $P_\mathrm{s}$ as functions of
  $\alpha$ for a black planet (gray, dashed--triple--dotted line), a
  white planet (blue, dashed--dotted line), and a planet with a
  gaseous atmosphere and its surface overed by alternating black and
  white pixels (black, solid line). The latter curves coincide with
  those calculated using horizontally homogeneous planets and a
  weighted sum approximation (red, dashed line). All planets are
  cloud--free.}
\label{fig:var_co_b_frst}
\end{figure}

Figure~\ref{fig:var_co_b_frst} shows $\pi F_\mathrm{n}$ and
$P_\mathrm{s}$ as functions of phase angle $\alpha$ for a planet with
its surface covered by alternating black and Lambertian reflecting
white pixels as calculated with the HI-code and the HH-code (the
latter combined with the weighted sum approximation, see
Eq.~\ref{eq:weightav}).  For comparison, $\pi F_\mathrm{n}$ and
$P_\mathrm{s}$ are also shown for horizontally homogeneous black and
white planets.

%-------------------------------------------------------------------------
% Appendix: Figure 3: made with MX_PL_PARAL/comp_pzl.pro
%-------------------------------------------------------------------------

The absolute difference between the fluxes of starlight reflected by
the black-and-white planet as calculated using the HI-code and the
HH-code is smaller than 7$\times 10^{-5}$, and between the degrees of
polarisation smaller than 0.005 across the whole phase angle range. It
appears that the differences between the results of the two codes
depend slightly on the surface albedo: decreasing the albedo of pixels
from 1.0 to 0.24 (which is typical for a sand surface at 0.55~$\mu$m),
the difference in the flux decreases to 5$\times 10^{-5}$ and in the
polarisation to 0.004.

The comparison with the HH-code applied to horizontally homogeneous
planets and using the weighted sum approximation shows that our
HI-code is accurate enough for application to horizontally
inhomogeneous planets, provided enough pixels are used across the
disk.

\end{appendices}

%\clearpage
%%%%%%%%%%%%%%%%%%%%%%%%%%%%%%%%%%%%%%%%%%%%%%%%%%%%%%%%%%%%%%%%%%%%%%%%%%
% Figures
%%%%%%%%%%%%%%%%%%%%%%%%%%%%%%%%%%%%%%%%%%%%%%%%%%%%%%%%%%%%%%%%%%%%%%%%%%

%%%%%%%%%%%%%%%%%%%%%%%%%%%%%%%%%%%%%%%%%%%%%%%%%%%%%%%%%%%%%%%%%%%%%%%%%%

\end{document}